\documentclass[conference]{IEEEtran}

\usepackage[T1]{fontenc}
\usepackage[utf8]{inputenc}
\usepackage{amsmath,amssymb,amsfonts}
\usepackage{graphicx}
\usepackage{caption}
\usepackage{booktabs}
\usepackage{microtype}
\usepackage[hidelinks]{hyperref}
\makeatletter
\g@addto@macro\UrlBreaks{\UrlOrds}
\makeatother
\usepackage{enumitem}
\setlist{nosep,leftmargin=1.4em}
\newcommand{\projectrepository}{\url{https://github.com/aminitech/aminiulap-digital-twin}}
\newcommand{\releasetag}{\texttt{v1.0-paper2}}
\newcommand{\releasedoi}{\texttt{10.5281/zenodo.21708211}}

\title{Sovereign Cognitive Digital Twins: Fusing 6G ISAC,
AI-RAN, and Zero-Trust Edge Grids for National Resilience in the Global South}

\author{%
\IEEEauthorblockN{Zoe Aiyanna M. Cayetano}
\IEEEauthorblockA{\textit{Amini}\\
Bridgetown, Barbados\\
zoe@amini.ai}
\and
\IEEEauthorblockN{George M. Gichuru}
\IEEEauthorblockA{\textit{Amini}\\
Bridgetown, Barbados\\
george@amini.ai}
\and
\IEEEauthorblockN{Taijuo T. Morris}
\IEEEauthorblockA{\textit{Amini}\\
Bridgetown, Barbados\\
taijuo.int@amini.ai}
}

\begin{document}
\maketitle

\begin{abstract}
Climate-hazard-prone archipelagic and small-island
developing states (SIDS) confront an existential risk profile---accelerating
sea-level rise, intensifying tropical cyclones, and storm surge---while
simultaneously suffering the sparse ground-based instrumentation that makes
timely hazard perception difficult. This paper argues for a paradigm shift from
passive cellular connectivity to the \emph{Network as a Sensor} (NaaS), realized
through a \emph{Sovereign Cognitive Digital Twin} (S-CDT): a federated national
digital twin whose perceptual substrate is the 6G radio interface itself. Rather
than deploying expensive, disjoint environmental sensing systems, a vulnerable
nation can reuse the Integrated Sensing and Communication (ISAC) waveforms of its
own network as a distributed radar mesh, feeding a closed-loop cognitive
orchestration engine for real-time state estimation, predictive hazard
forecasting, and automated mitigation. We ground the architecture in the
federation lessons of existing national and urban digital twin programs,
Virtual Singapore and Destination Earth, and in the Gemini Principles governance
framework; we specify a six-layer S-CDT stack in which ISAC structurally collapses the
boundary between the dynamic-data and communication layers; we map the physical
layer to the ETSI GR ISC~001 and 3GPP Release~19 sensing frameworks; and we
formulate a belief-state control loop---an Extended Kalman Filter feeding a
Proximal Policy Optimization agent---designed to absorb O-RAN telemetry delay.
The control loop is specified but not evaluated here; it is a design
contribution, not a deployed or simulated result. Beyond the reference
architecture, we implement a reproducible, CPU-only geodata-to-ray-tracing
pipeline over a 2~km study area at the Barbados Heritage District, Newton
Plantation:
576 LiDAR-height buildings, a 70$\times$70 terrain grid, and the government tower
register are transformed through Blender into a Sionna RT scene. On identical
geometry, median best-server path gain decreases from $-106$~dB at 1.8~GHz to
$-122$~dB at 10~GHz, and concrete-only 28/60~GHz runs yield $-130/-136$~dB with
coverage contracting to line-of-sight lobes. Ground-following maps additionally
resolve ridge shadowing that a horizontal receiver plane cannot represent, though we
quantify this effect at under $1$\,\% of the study area and correct an earlier
receiver-plane discretisation that overstated it.
These uncalibrated, $1\!\times\!1$ V-polarized simulations are explicitly framed
as a \emph{towards-6G} site model of $\mathbf{H}_{\mathrm{background}}$, not as an
operational ISAC deployment. Finally, we treat data sovereignty and
physical-layer zero-trust security
as first-order design constraints, proposing subsystem disaggregation,
cross-layer anomaly detection, and privacy-tiered waveforms. Barbados
($166~\mathrm{km}^2$, $\sim$280k population) is used as a tractable reference
deployment, with the Philippines as an archipelagic generalization.
\end{abstract}


\section{Introduction}

\subsection{The Climate Existential Threat}
Archipelagic and small-island developing states occupy the sharp edge of
anthropogenic climate change. For nations such as Barbados in the Caribbean and
the Philippines in the western Pacific, sea-level rise, coastal erosion, and the
intensification of typhoons and storm surge are not distant projections but
present-tense determinants of national survival, economic continuity, and
population safety~\cite{undp_sids,wef_barbados}. These same nations are, in
general, the most weakly instrumented: dense networks of tide gauges, weather
radar, and hydrological sensors are capital-intensive to deploy and, critically,
fragile precisely during the extreme events for which they are most needed.
The result is a perception gap---decisions with hours-to-minutes stakes made on
data with hours-to-days latency. Our prior work addressed the complementary
problem of keeping the bearer network itself alive through such events, via
database-free TV white space sensing for disaster-resilient connectivity in
SIDS~\cite{sidsense}; the present paper asks what that network, once it exists,
can be made to \emph{perceive}.

\subsection{A Paradigm Shift: The Network as a Sensor}
We propose closing that gap not by multiplying dedicated sensors but by
re-conceiving infrastructure the nation is already motivated to build. The
transition from fifth- to sixth-generation mobile networks introduces
\emph{Integrated Sensing and Communication} (ISAC), in which the radio interface
that carries data simultaneously performs radar-like sensing of its
environment~\cite{ericsson_isac,etsi_isc001}. Under this \emph{Network as a
Sensor} (NaaS) paradigm, every base station is also an environmental
instrument, and the coverage footprint of the network becomes the sensing
footprint of the nation.

\subsection{Thesis and Contributions}
Our thesis is that vulnerable nations can deploy a unified \emph{Sovereign
Cognitive Digital Twin} (S-CDT) that uses the 6G radio interface as a
continuous perceptual nervous system, providing real-time modeling, predictive
hazard forecasting, and closed-loop disaster mitigation, under national
sovereign control of models and data. This paper contributes: (i) a federated
S-CDT reference architecture aligned with established national-digital-twin
governance (\S\ref{sec:baseline}); (ii) a six-layer stack in which ISAC
collapses the dynamic-data and communication layers (\S\ref{sec:stack}); (iii)
a standards-grounded treatment of the physical sensing layer
(\S\ref{sec:physical}); (iv) the integration of an established belief-state
control pattern~\cite{tiwari_belief} into a sovereign twin, formulated to
tolerate RAN telemetry delay and constrained to on-territory execution
(\S\ref{sec:cognitive}); (v)
a sovereign, physical-layer zero-trust security and privacy design that composes
published multi-domain-security~\cite{keskin_multidomain} and
ISAC-privacy~\cite{gunlu_privacy} primitives into a national access-control
boundary (\S\ref{sec:security});
and (vi) a reproducible CPU-only, authoritative-geodata-to-Sionna pipeline with
quantified simulation results that upgrades the deployed RF layer from stochastic
national planning to deterministic site studies (\S\ref{sec:rt-method}).

\noindent We are explicit about where the novelty in this list does and does not
lie. Contributions (iv) and (v) claim no new estimator, policy, or cryptographic
primitive: the belief-state formulation is due to Tiwari
\emph{et al.}~\cite{tiwari_belief}, the privacy-level taxonomy to G\"unl\"u
\emph{et al.}~\cite{gunlu_privacy}, the multi-domain security framing to Keskin
\emph{et al.}~\cite{keskin_multidomain}, and the compression theory that
\S\ref{sec:cognitive} defers to, to Pan \emph{et al.}~\cite{pan_compression}.
What is ours in (iv) and (v) is the composition---placing those mechanisms under
a single national sovereignty boundary---and that composition is specified here,
not evaluated. The contributions that rest on work performed for this paper are
(i), (ii), (vi), and the sovereignty argument that motivates them.

\section{Grounding the Baseline: The National Digital Twin Paradigm}
\label{sec:baseline}

\subsection{The Federated Architecture}
The single most consequential design decision is to reject the monolith. The UK
National Digital Twin Programme's central, hard-won lesson is that a national
twin is not one model of everything but a \emph{federated ecosystem} of
domain-specific twins, connected through shared identifiers, open standards, and
a common trust framework~\cite{ndtp,gemini}. Federation is what makes the system
buildable incrementally, governable across institutional owners, and resilient
to the failure or replacement of any single component.

\subsection{Governance via the Gemini Principles}
We structure S-CDT governance around the nine Gemini
Principles~\cite{gemini}, grouped as:
\begin{itemize}
\item \textbf{Purpose}---public good, actionable insight, and measurable
economic and climatic value.
\item \textbf{Trust}---security by default (here elevated to \emph{zero trust}),
transparent data provenance, and high-quality validation.
\item \textbf{Function}---federated interoperability, scalability, and long-term
evolutionary capability.
\end{itemize}
For a sovereign deployment the ``trust'' cluster is load-bearing: the twin's
outputs must be defensible enough to justify evacuations and to underwrite
climate finance, which demands auditable provenance rather than best-effort
data hygiene.

\subsection{The Design Brackets}
Two operational national/urban twins bracket the design space. The
\emph{built-environment bracket} is Virtual Singapore (Dassault Syst\`emes with
GovTech, SLA, and NRF): a semantically rich 3D model fused with live smart-nation
sensor feeds, driving micro-mobility, solar-potential, and flood
simulation~\cite{vsg}. The \emph{earth-system bracket} is Destination Earth
(DestinE) of the European Commission, with ESA, EUMETSAT, and ECMWF: planetary
climate-adaptation and extreme-events twins running on high-performance-computing
architectures over a multi-source data lake~\cite{destine}. The S-CDT merges the
brackets---island-scale built-environment fidelity coupled to earth-system
climate dynamics---and adds the 6G shift that neither exemplar yet exploits:
\emph{the connecting network is itself the primary real-time sensor}, redefining
the boundary of ``live data.''

\section{The Unified S-CDT Layered Stack and the ISAC Transformation}
\label{sec:stack}

\subsection{The Core Stack}
The S-CDT is organized as six layers over a vertical trust-and-governance spine:

\begin{enumerate}
\item \textbf{Geospatial Ground Truth}---high-resolution terrain, bathymetry,
building envelopes, and critical utility networks. Barbados's
$166~\mathrm{km}^2$ area and $\sim$280k population make complete, high-fidelity
national modeling extraordinarily tractable relative to a large state.
\item \textbf{Systems of Record}---dynamic registries of land, business,
population, and infrastructure assets.
\item \textbf{Dynamic Data / Nervous System}---fusion of satellite Earth
Observation, IoT telemetry, and the 6G ISAC radio-sensing network.
\item \textbf{Simulation and Surrogates}---physics-based hydrology and coastal
hydrodynamic models, statistical agent-based evacuation models, and deep-learning
surrogate models for real-time ray-tracing and wave propagation.
\item \textbf{Synchronization Engine}---reconstruction of current-time state
estimates from dynamic physical telemetry.
\item \textbf{Interaction and Orchestration}---generative-AI natural-language
query interfaces, executive dashboards, and automated control-API pipelines.
\end{enumerate}

\subsection{The ISAC Structural Fusion}
In the classical layered view, the communication infrastructure is plumbing that
\emph{transports} data produced by a physically separate sensing layer. ISAC
dissolves this separation: the physical radio waves of the network double as a
distributed radar mesh, so the Dynamic-Data Layer and the communication
infrastructure become one substrate~\cite{ericsson_isc001,etsi_isc001}. Sensor
deployment as a distinct capital program becomes largely obsolete; instead,
sensing is a software-and-spectrum function of the network the nation deploys
for connectivity. For an operator-integrator this collapses three roles into a
single asset---the network is simultaneously (a)~physical infrastructure the twin
\emph{models}, (b)~the backbone that \emph{carries} twin data, and (c)~the
distributed sensor that \emph{feeds} the twin.

\section{The Network as the Perceptive Nervous System: Physical Layer and Standards}
\label{sec:physical}

\subsection{Dual-Function Radar-Communication}
At the physical layer, ISAC is realized as Dual-Function Radar-Communication
(DFRC): a 6G base station uses shared spectrum and shared hardware to transmit a
waveform that simultaneously delivers high-throughput data and yields
radar-grade observables---range, angle (angle-of-arrival), radial velocity
(Doppler), and micro-structure---by processing echoes and multipath. The ISAC
propagation channel is conveniently decomposed as
\begin{equation}
\mathbf{H}_{\mathrm{ISAC}} = \mathbf{H}_{\mathrm{target}} +
\mathbf{H}_{\mathrm{background}},
\end{equation}
where $\mathbf{H}_{\mathrm{target}}$ captures returns from dynamic sensing
targets and $\mathbf{H}_{\mathrm{background}}$ models the quasi-static
environment~\cite{tr38901,rel19survey}. Estimating and subtracting a learned
background is precisely what lets the network isolate a moving hazard---or a
weather cell---from clutter. This additive form is a deliberate simplification
for architectural exposition, and we use it only as such. The Release~19 model is
substantially richer: the target term is built from concatenated sub-channels
over monostatic and bistatic radar-cross-section scattering points, and the
background term from a reference-point-based geometry-based stochastic model
carrying its own angular and delay spreads~\cite{rel19survey}. Nothing in the
architecture below depends on the simplified form; a Release~19-conformant
implementation would substitute the full extended-GBSM formulation without
structural change.

\subsection{Standardization Milestones}
\textbf{ETSI.} The ETSI ISG ISAC group's first report, GR ISC~001 (April 2025),
establishes 18 advanced use cases, three integration levels (tight,
intermediate, and loose), and six sensing modes for 6G
sensing~\cite{etsi_isc001}; subsequent reports (e.g.\ GR ISC~003, 2026) address
system and RAN architectures~\cite{etsi_isc003}. \emph{(We note a common
mis-citation: the 18-use-case report is GR ISC~001, not GR ISC~004.)} The six
modes are defined by transmitter/receiver placement:
\begin{itemize}
\item \textbf{Monostatic}: TRP-based (Tx/Rx co-located at a
transmission-reception point) and UE-based.
\item \textbf{Bistatic}: TRP$\rightarrow$TRP, TRP$\rightarrow$UE,
UE$\rightarrow$TRP, and UE$\rightarrow$UE.
\end{itemize}
Where this paper refers to ``the six sensing modes,'' it is this ETSI taxonomy
that is meant.

\noindent\textbf{3GPP.} Release~19 (study concluded May 2025) specifies wireless
sensing use cases in TR~22.837 and extends the TR~38.901 channel model for
ISAC~\cite{tr22837,tr38901}. Its channel-model work does not adopt the ETSI
six-mode taxonomy. It instead categorizes ISAC transmitters and receivers into
four node types---TRP, normal UT, vehicle UT, and aerial UT---and enumerates the
nine propagation links these pairings generate: TRP--TRP, TRP--normal~UT,
TRP--vehicle~UT, TRP--aerial~UT, normal~UT--normal~UT, normal~UT--vehicle~UT,
normal~UT--aerial~UT, vehicle~UT--vehicle~UT, and
aerial~UT--aerial~UT~\cite{rel19survey}. The two enumerations are complementary
rather than competing: the ETSI modes classify sensing geometry by where the
transmitter and receiver sit relative to one another, while the 3GPP links
enumerate the node-type pairings a channel model must cover. Sensing spans FR1,
the emerging FR3 mid-band, and FR2 mmWave, trading coverage against
range/velocity resolution. Release~20 continues the architecture study.

\subsection{Climbing the Sensing Maturity Ladder}
Each environmental sub-twin climbs the maturity ladder
independently---\emph{Descriptive} $\rightarrow$ \emph{Diagnostic} $\rightarrow$
\emph{Predictive} $\rightarrow$ \emph{Prescriptive} $\rightarrow$
\emph{Autonomous}---with the predictive-and-above rungs driven natively by edge
AI and network sensing. Three exemplar capabilities:

\paragraph{Moving-object detection.} Coordinated, multi-static tracking of
non-cooperative aerial hazards (unregistered drones over critical
infrastructure) and prioritization of emergency vehicles, using Doppler
signatures without requiring the target to carry a device.

\paragraph{Environmental and weather mapping.} Hyper-local rainfall estimation
inferred from frequency-specific attenuation. Specific attenuation follows the
ITU-R power law
\begin{equation}
\gamma_R = k\,R^{\alpha} \quad [\mathrm{dB/km}],
\end{equation}
with rain rate $R$ and band-dependent coefficients $k,\alpha$~\cite{itur838};
the path integral of $\gamma_R$ is observable as excess attenuation in the
channel state, so mmWave/FR3 links act as a dense mesh of virtual rain
gauges~\cite{raingauge}, enabling flash-flood and track-washout prediction well
below the resolution of sparse ground radar.

\paragraph{Radio-environment analysis.} Continuous RF fingerprinting and
multipath-delay-profile indexing dynamically update the local clutter map that
constitutes $\mathbf{H}_{\mathrm{background}}$, both improving sensing and
providing a change-detection channel in its own right. The Newton implementation
in \S\ref{sec:rt-method} instantiates the static, site-specific baseline for this
quantity: deterministic geometry, material labels, terrain, tower locations,
and resolved paths become the background model against which future CSI/ISAC
observations can detect change.

\subsection{Path Loss and Material-Aware Attenuation}
In Sionna RT, path loss in the SCOPE model uses free-space path loss (FSPL) as a
baseline for each individual ray to determine its attenuation, and combines this
with material reflections to calculate total path loss:
\begin{equation}
PL_{\mathrm{total}} = FSPL + L_{\mathrm{material}} + L_{\mathrm{other}}
\end{equation}
where $PL_{\mathrm{total}}$ is the total path loss, $L_{\mathrm{material}}$ is
the loss introduced by the ray interacting with a material, and
$L_{\mathrm{other}}$ is any additional loss the ray experiences. All terms are in
dB. FSPL is
calculated as~\cite{rappaport}:
\begin{equation}
FSPL_{\mathrm{dB}} = 20\log_{10}(d) + 20\log_{10}(f) + 32.44
\end{equation}
where $d$ is the distance in kilometers and $f$ is the frequency in MHz.

\paragraph{MAPL and model connection.} While total path loss captures the loss
incurred over a link, whether that link is actually usable depends on the
system's tolerance for loss---captured by the maximum allowable path loss
(MAPL). A link remains viable only where $PL_{\mathrm{total}} \leq MAPL$; beyond
this threshold, connectivity between UEs fails due to insufficient signal
strength. For the distributed radar mesh, this threshold effectively sets a
maximum node spacing and determines where coverage gaps emerge.

\section{Closed-Loop Cognitive Orchestration: AI-RAN and Belief-State Control}
\label{sec:cognitive}

\noindent\emph{Scope note.} This section specifies the cognitive layer of the
reference architecture. None of it is deployed: the belief-state estimator, the
reinforcement-learned orchestrator, and the edge compression stage are roadmap
Phases~3 and~4 (\S\ref{sec:roadmap}), and the equations below are design
formulations that have not been evaluated on the Barbados deployment. Section
\ref{sec:impl} reports what is actually running.

\subsection{AI-RAN Convergence}
Perception without timely action is merely telemetry. The design distributes
intelligence into the Radio Access Network itself: model training via federated
learning across sites (keeping raw observations local), and inference at the
O-RAN Near-Real-Time and Non-Real-Time RAN Intelligent Controllers
(RICs)~\cite{oran_isac}. The Non-RT RIC would host the digital-twin models and
policy learning; the Near-RT RIC would execute low-latency control (beam and
resource allocation) as an xApp.

\subsection{Belief-State Control under Telemetry Delay}
The central control obstacle is delay: O-RAN telemetry can reach the controller
tens of milliseconds late (up to $\sim$100~ms), so the instantaneous
measurement is a stale view of a fast-moving hazard field. The design therefore
controls on a \emph{belief state} rather than raw observations.

The belief-state pattern for this problem is established prior work, and we
adopt it rather than propose it. Tiwari \emph{et al.} give a digital-twin-assisted
belief-state reinforcement-learning formulation for latency-robust ISAC in the
same ${\sim}100$~ms O-RAN telemetry-delay regime, pairing a filtered state
estimate with a learned policy, and---unlike this paper---report a quantitative
evaluation of it~\cite{tiwari_belief}. What we set out below is that formulation
placed inside a sovereign twin: the estimator and the policy are constrained to
execute on on-territory compute (\S\ref{sec:security}), the observation vector is
a national hazard field rather than a vehicular or cellular scene, and the reward
carries an explicit sovereignty and security term. The formulation itself is
theirs. An Extended Kalman Filter (EKF) hosted inside the twin would maintain an
estimate of the latent environmental and network state $\mathbf{x}_t$. The prediction step is
\begin{align}
\hat{\mathbf{x}}_{t|t-1} &= f(\hat{\mathbf{x}}_{t-1|t-1},\mathbf{u}_t),\\
\mathbf{P}_{t|t-1} &= \mathbf{F}_t \mathbf{P}_{t-1|t-1}\mathbf{F}_t^{\top} +
\mathbf{Q}_t,
\end{align}
with $\mathbf{F}_t = \left.\partial f/\partial\mathbf{x}
\right|_{\hat{\mathbf{x}}_{t-1|t-1}}$. To absorb a measurement delay $\tau$, the
update fuses the delayed observation $\mathbf{z}_{t-\tau}$ against the
correspondingly retrodicted state,
\begin{align}
\mathbf{K}_t &= \mathbf{P}_{t|t-1}\mathbf{H}_t^{\top}
\!\left(\mathbf{H}_t \mathbf{P}_{t|t-1}\mathbf{H}_t^{\top}
+ \mathbf{R}_t\right)^{-1},\\
\hat{\mathbf{x}}_{t|t} &= \hat{\mathbf{x}}_{t|t-1}
+ \mathbf{K}_t\!\left(\mathbf{z}_{t-\tau}
- h(\hat{\mathbf{x}}_{t-\tau|t-1})\right).
\end{align}
The resulting belief $s_t = (\hat{\mathbf{x}}_{t|t}, \mathbf{P}_{t|t})$---the
mean estimate together with its covariance---is a sufficient statistic that
explicitly carries the controller's uncertainty into the decision.

\subsection{Reinforcement-Learned Orchestration}
A Proximal Policy Optimization (PPO) agent would map the belief state to control
actions $a_t$---beamforming power, sensing dwell/scan allocation, and mode
selection across the six ISAC modes---maximizing the clipped surrogate
objective
\begin{equation}
L^{\mathrm{CLIP}}(\theta) = \mathbb{E}_t\!\left[\min\!\big(\rho_t(\theta)\hat{A}_t,\;
\mathrm{clip}(\rho_t(\theta),1-\epsilon,1+\epsilon)\hat{A}_t\big)\right],
\end{equation}
with probability ratio $\rho_t(\theta)=\pi_\theta(a_t\mid s_t)/
\pi_{\theta_{\mathrm{old}}}(a_t\mid s_t)$ and advantage estimate $\hat{A}_t$. The
reward balances the competing objectives of a shared sensing/communication
resource,
\begin{equation}
r_t = \alpha\,\mathcal{A}_{\mathrm{sens}}
+ \beta\,\mathcal{T}_{\mathrm{comm}}
- \gamma\,\mathcal{E}
- \delta\,\mathcal{R}_{\mathrm{sec}},
\end{equation}
rewarding sensing accuracy $\mathcal{A}_{\mathrm{sens}}$ and communication
throughput $\mathcal{T}_{\mathrm{comm}}$ while penalizing energy $\mathcal{E}$
and residual security risk $\mathcal{R}_{\mathrm{sec}}$. The intended behavior
is that under an active-hazard regime the learned policy shifts resource toward
tracking, while under nominal conditions it favors throughput. No policy has
been trained; verifying that this reward shaping induces the intended behavior
is future work.

\subsection{Edge Observation Compression}
Multi-static sensing produces high-dimensional 3D point clouds that would
saturate rate-limited emergency backhaul exactly when links are degraded. The
design applies an autoencoder (AE) at the gNodeB for dimension reduction: an encoder
$f_\phi$ maps a point cloud $\mathbf{P}$ to a compact code
$\mathbf{z}=f_\phi(\mathbf{P})$, transmitted in place of the raw cloud and
reconstructed by $g_\psi$ at the core, trained to minimize
\begin{equation}
\mathcal{L}_{\mathrm{AE}} = d_{\mathrm{CD}}\!\big(\mathbf{P},
g_\psi(f_\phi(\mathbf{P}))\big) + \lambda\lVert\mathbf{z}\rVert_1,
\end{equation}
where $d_{\mathrm{CD}}$ is the Chamfer distance and the $\ell_1$ term encourages
a sparse, low-rate code. The intent is to preserve hazard-relevant geometry
under a strict backhaul budget; the achievable rate-distortion trade-off is left
to future work. We note that the theory this defers to already exists: Pan
\emph{et al.} characterize observation compression in rate-limited closed-loop
distributed ISAC systems, carrying the analysis from signal reconstruction
through to control performance~\cite{pan_compression}. Their results, rather than
a fresh derivation, are the appropriate baseline against which any S-CDT
compression stage should be evaluated.

\section{Sovereign Data Protection and Physical-Layer Zero Trust}
\label{sec:security}

\subsection{The Sovereign Compute Grid}
A twin that governs evacuations and underwrites climate finance is national
critical infrastructure. Climate-vulnerable nations must therefore retain
sovereign control of AI weights, models, and spatial telemetry rather than
renting perception from external cloud monopolies. The S-CDT is designed for an
on-territory sovereign compute grid---edge inference at the RAN, national core
for training and archival---so that both the raw sensing of the population and
the learned models derived from it remain under domestic jurisdiction. This is
also a resilience property: sovereign edge autonomy keeps the twin operating
when international connectivity is severed by the very disaster it must manage.
The CPU-only propagation workflow in \S\ref{sec:rt-method} demonstrates this
principle concretely: its Mitsuba LLVM backend requires neither CUDA nor an
external cloud service, while a GPU runbook remains an optional scale-up path.

\subsection{Zero-Trust Subsystem Disaggregation}
Because a network that senses the nation makes its own compromise a
national-security event rather than a service outage, the design adopts a
disaggregated trust model in the spirit of OPPO's ``Minimized Kernel~$+~N$
Subsystems'' 6G architecture~\cite{oppo_kernel,oppo_sec}: a minimal, formally
hardened kernel would provide base connectivity, while sensitive
functions---national-security sensing, emergency services, and the
environmental-sensor plane---would run as isolated subsystems with independent
trust domains. The intended property is that no plane implicitly trusts another,
so that the environmental-sensing plane cannot be pivoted into from a
compromised consumer-data plane. No such subsystem isolation exists in the
current deployment, which runs as a single trust domain;
\S\ref{sec:impl-governance} reports what is enforced today.

\subsection{Cross-Layer Anomaly Detection}
Physical-layer observables are themselves a security sensor. The general case for
this is made by Keskin \emph{et al.}, who set out a multi-domain security
framework for 6G ISAC spanning the cyber-physical, physical-layer, and protocol
domains, and argue that cryptographic protocol mechanisms alone cannot detect
lower-layer attacks~\cite{keskin_multidomain}. We follow that framing and apply
it to a national environmental-sensing plane rather than to their transportation
setting: fusing angle-of-arrival (AoA), Doppler shift, and channel state
information (CSI) with higher-layer protocol state to detect physical-layer
attacks that are invisible above the PHY---signal spoofing, GNSS/GPS jamming, and
tampering with Reconfigurable Intelligent Surfaces (RIS). A spoofed transmitter
that is protocol-correct but geometrically impossible (inconsistent AoA/Doppler
for its claimed identity) would be flagged by such a cross-layer detector. No
detector of this kind is implemented in the current deployment.

\subsection{Privacy-by-Design Waveforms}
NaaS sensing is powerful and politically sensitive: the same channel that maps
topography can, at fine resolution, read human behavior. We adopt the three-level
classification of privacy-sensitive ISAC data introduced by G\"unl\"u
\emph{et al.}~\cite{gunlu_privacy}, who organize it into location and environment
data, behavioral data, and physiological data, and survey the corresponding
mitigations. Using their levels as our privacy tiers---\textbf{L1 Environmental}
(terrain, hydrology, weather), \textbf{L2 Behavioral} (aggregate mobility,
presence), and \textbf{L3 Physiological} (fine micro-Doppler, e.g.\
respiration/gait)---we attach escalating consent and access controls to each and
bind them to the sovereignty boundary of \S\ref{sec:security}. The taxonomy is
theirs; what we add is its use as an access-control boundary inside a
nationally-governed twin. To enforce the boundary at the physical layer, the
design injects calibrated noise into the reported CSI,
\begin{equation}
\tilde{\mathbf{H}} = \mathbf{H} + \mathbf{N},\qquad
\mathbf{N}\sim\mathcal{P}\big(\sigma_{\mathrm{beh}}\big),
\end{equation}
where $\mathcal{P}$ is designed to destroy the high-frequency micro-Doppler
components that carry behavioral/physiological signatures while preserving the
low-frequency structural returns needed for topography and weather sensing. The
design goal is a network that delivers high-resolution environmental perception
at L1 while being unable to export L2/L3 signatures without explicit,
tier-appropriate authorization. We state this as a design target rather than a
guarantee: establishing the leakage bound for a given $\mathcal{P}$, and the
resulting privacy-versus-sensing-utility trade-off curve, is future work
(\S\ref{sec:roadmap}). In the current deployment only L1 data is served, so no
L2/L3 export path yet exists to gate.

\section{Deployment Blueprint: Barbados and the Archipelagic Generalization}
\label{sec:deploy}

We propose a \emph{minimum viable twin} that proves the full vertical slice on a
single existential question before federating outward. Its components: (1) a
national basemap bootstrapped from Earth-observation foundation models; (2) at
least one network-derived live feed---microwave-link/CSI rainfall---designed
ISAC-ready, alongside a tide/weather feed; (3) one predictive model (coastal
inundation under a parametric storm scenario) driven by the live rainfall field;
(4) a generative natural-language query interface for decision-makers; and (5)
node and sensor attestation plus data provenance on a sovereign ledger to make
outputs finance- and evacuation-grade. Federation then proceeds domain by
domain---economy and tourism next, coupled to the climate twin through
network-derived mobility---while every newly deployed base station is specified
ISAC-ready so the sensing mesh densifies as a byproduct of connectivity
roll-out. Barbados serves as a bounded, high-fidelity reference; the Philippines
generalizes the design to a multi-thousand-island archipelago where inter-island
non-terrestrial (satellite) ISAC and edge autonomy become decisive.

\begin{figure*}[t]
\centering
\includegraphics[width=0.62\textwidth]{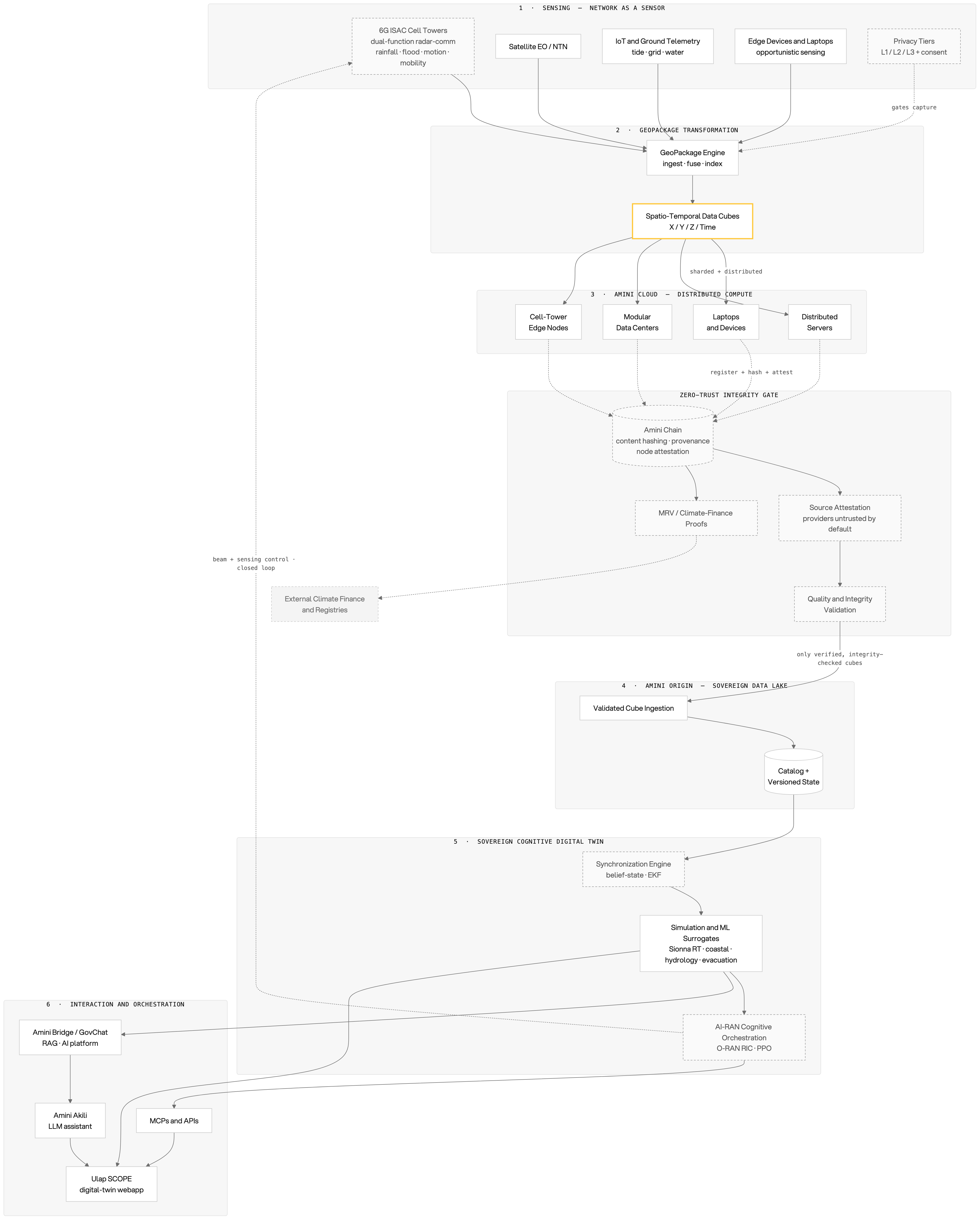}
\caption{The S-CDT reference architecture: a six-layer stack over a vertical
zero-trust and integrity spine. \emph{This figure depicts the target
architecture, not the current deployment.} In the target design, sensing
(Network-as-a-Sensor) would feed the Amini GeoPackage engine producing
spatio-temporal data cubes; cubes would be sharded across the Amini Cloud
(decentralized compute infrastructure) node
network and anchored on Amini Chain (a private, off-chain ledger for content
hashing, provenance, and node attestation); a zero-trust integrity gate would
admit only verified cubes to the Amini Origin national sovereign data lake
(termed ``Bajan-X'' in the Barbados deployment); and orchestration would close
the loop back to the sensing layer. Of these elements, the ingest, cube-store,
catalog, and interaction paths are built; the ISAC feed, the ledger anchoring,
the cryptographic admission gate, the belief-state synchronization engine, and
the AI-RAN control loop are not yet implemented. Table~\ref{tab:map} and
Table~\ref{tab:gov} state the built/designed split precisely, and
\S\ref{sec:roadmap} gives the phasing.}
\label{fig:arch}
\end{figure*}

\section{Barbados Implementation and Towards-6G Propagation Results}
\label{sec:impl}
The preceding sections describe the target S-CDT. This section reports what has
actually been built to date on the Barbados deployment, inside the Ulap SCOPE
application (a connectivity network planner built on Sionna RT).

\noindent\emph{Terminology.} Two distinct systems in this work are both properly
called digital twins, and we name them separately throughout. The
\textbf{national twin} is the infrastructure and hazard twin built by
\texttt{twin\_ingest} and served by \texttt{twin\_service} over
\texttt{/api/v1/twin/*}: authoritative government layers, buildings, utilities,
and vulnerability grids. The \textbf{RF scene twin} is the deterministic
ray-tracing scene twin: Blender geometry solved in Sionna RT to produce
propagation predictions. They share the same sovereign source contract and the
same provenance chain, but they are different artifacts with different
maturity levels, and a claim about one is not a claim about the other.

Measured against the sensing-maturity ladder of
\S\ref{sec:physical}, the national twin sits at the
\emph{descriptive/diagnostic} rungs: it renders the nation and its exposure and
joins hazard return-periods to assets. The Newton propagation component now
adds an uncalibrated \emph{predictive simulation} rung through deterministic ray
tracing, but the deployment does not yet ingest ISAC measurements or run the
EKF/PPO cognitive loop.
Figures~\ref{fig:rf-coverage}--\ref{fig:island-overview} provide compact visual
evidence of the implemented planning, provenance, inspection, interaction, and
multi-source fusion workflows; Figures~\ref{fig:freq-sweep}--\ref{fig:multitower}
report the deterministic propagation outputs.

\subsection{Mapping the deployment to the reference stack}
Figure~\ref{fig:arch} shows the end-to-end sovereign dataflow;
Table~\ref{tab:map} maps the conceptual six-layer stack of
\S\ref{sec:stack} to the deployed system. The build realizes geospatial ground
truth, systems of record, conventional dynamic feeds, deterministic RF
simulation, and the interaction surface. Belief-state synchronization, active
ISAC sensing, ledgered trust, and closed-loop orchestration remain scheduled
(\S\ref{sec:roadmap}).

\begin{table*}[t]\footnotesize\centering
\caption{Reference-stack layers mapped to the Ulap SCOPE deployment.
[B]~=~built, [D]~=~design.}\label{tab:map}
\begin{tabular}{@{}p{2.5cm}p{11.1cm}p{2.0cm}@{}}
\toprule
S-CDT layer & Deployed in Ulap SCOPE today & State\\
\midrule
1. Geospatial ground truth & Authoritative buildings, terrain/elevation, utilities,
hazards, vulnerability grids, and tower inventory; national web layers plus
EPSG:21292 propagation scenes. & [B]\\
2. Systems of record & \texttt{twin\_ingest} builds the GeoPackage/PMTiles
store and asset registries; \texttt{catalog.json} and
\texttt{scene\_manifest.json} carry source and transformation metadata. & [B]\\
3. Dynamic data / nervous system & 3D-PAWS/CHORDS weather, OpenSky, CelesTrak,
Sentinel/Copernicus, Open-Meteo and OpenCellID connectors. ISAC radar/CSI sensing
is not yet a feed. & [B] conventional; [D] ISAC\\
4. Simulation and surrogates & Island-wide stochastic planning plus deterministic
Sionna RT site studies, terrain-following maps, frequency/material sweeps,
interference, transect and mobility solves. & [B] RF; [D] calibrated climate/ISAC\\
5. Synchronization engine & Live layers are hydrated independently; no EKF
belief-state estimator, uncertainty propagation, or cross-domain temporal state
history yet. & [D]\\
6. Interaction/orchestration & SCOPE 3D map, KPIs, inspect/hazard dashboards,
Amini Akili (LLM assistance) and live UE-Sim ray-traced snapshots; no automated
control loop. & [B]
interaction; [D] orchestration\\
Zero-trust spine & Per-output source attribution, graceful degradation,
structural/physics validity checks and CPU-only sovereign execution; no signed
hash ledger or node attestation yet. & [B] partial; [D] cryptographic gate\\
\bottomrule
\end{tabular}
\end{table*}

\subsection{Data provenance: authoritative layers}
\label{sec:impl-provenance}
All 19 static layers derive from a single authoritative source, the Barbados
Geoportal of the Lands \& Surveys Department, received as 19 ESRI shapefiles
constituting a national infrastructure, hazard and vulnerability
database~\cite{bbdgeo}. Two facts govern every transformation. First, source
geometry is EPSG:21292 (Barbados 1938 / British West Indies Grid) and is reprojected
to EPSG:4326 for web mapping, while propagation scenes deliberately remain in
EPSG:21292 so ray lengths and terrain elevations are expressed in true meters; a
correctness invariant is that the building bounding
box must reproject to the whole island (verified at $[-59.650,\,13.045]$ to
$[-59.422,\,13.335]$). Second, critical-asset layers carry hazard return-period
fields (rainfall and coastal flood, landslide, seismic), which is what lets the twin
join assets to the climate-risk models. Table~\ref{tab:layers} is the ingested
inventory. The government \texttt{vulnerability\_*} grid is treated as authoritative
and supersedes the earlier census-derived parish proxy; the service degrades to the
lower-provenance proxy only when the grid is absent, and carries the source label
either way. For the Newton radio study, this provenance chain extends through the
physics: each simulated cell is tied to a national-grid coordinate, LiDAR-derived
\texttt{AVG\_HEIGHT}/\texttt{AVG\_DTM} attributes, the August~2023 national
antenna register, and an explicit propagation-material label.

\begin{table*}[t]\scriptsize\centering
\caption{Ingested static layer inventory (real counts from \texttt{catalog.json}).
Source: Barbados Geoportal, Lands \& Surveys Dept.; CRS EPSG:21292~$\rightarrow$~EPSG:4326.
\texttt{track}~=~serving strategy.}\label{tab:layers}
\begin{tabular}{@{}p{3.2cm}llrl@{\hskip 10pt}p{6.6cm}@{}}
\toprule
Layer (twin name) & Geom & Count & Track & Twin role\\
\midrule
\texttt{buildings} & polygon & 130{,}248 & tiles & Island-wide 3D on terrain\\
\texttt{roads} & line & 705 & geojson & Road network\\
\texttt{bridges} & line & 25 & geojson & Bridges (hazard-annotated)\\
\texttt{ports} & polygon & 4 & geojson & Ports \& BGI airport hotspots\\
\texttt{antennas} & point & 100 & geojson & Telecom / RF scene twin\\
\texttt{drinking\_water\_network} & line & 17{,}802 & tiles & Potable mains\\
\texttt{water\_mains} & line & 17{,}966 & tiles & Water mains\\
\texttt{wastewater\_plants} & point & 48 & geojson & Wastewater plants\\
\texttt{desalination\_plants} & point & 5 & geojson & Desalination (critical)\\
\texttt{dams} & point & 39 & geojson & Dams\\
\texttt{reservoirs} & point & 34 & geojson & Reservoirs\\
\texttt{pumping\_stations} & point & 19 & geojson & Pumping stations\\
\texttt{communal\_wells} & point & 38 & geojson & Wells\\
\texttt{individual\_wells} & point & 40 & geojson & Wells\\
\texttt{major\_projects} & point & 111 & geojson & Capital projects\\
\texttt{vulnerability\_global} & polygon & 13{,}029 & tiles & Authoritative gov.\ vulnerability grid\\
\texttt{vulnerability\_population} & polygon & 13{,}029 & tiles & Population exposure\\
\texttt{vulnerability\_environmental} & polygon & 13{,}029 & tiles & Environmental exposure\\
\texttt{vulnerability\_equipment} & polygon & 13{,}029 & tiles & Equipment / economic exposure\\
\bottomrule
\end{tabular}
\end{table*}

\subsection{Live feeds}
Beyond the static export, typed connectors hydrate live layers, one documented
ingress per source: 3D-PAWS weather stations via CHORDS, Open-Meteo, OpenSky
flights, CelesTrak TLEs (NTN constellations), Sentinel-2 L2A and Copernicus DEM via
the CDSE STAC, ESA WorldCover, Microsoft Global Building Footprints (fallback only),
Hansen forest change, Geofabrik/OSM, OpenCellID, TeleGeography submarine cables, and
PeeringDB. A strict provenance rule governs overlap: the government export always
wins where both exist (for example, government building footprints over Microsoft's),
and live feeds serve enrichment, hotspots and layers the export does not cover.

\subsection{Zero-trust governance: target versus enforced}
\label{sec:impl-governance}
The governance model follows \S\ref{sec:security} and the Gemini Principles: sources
untrusted by default, provenance mandatory rather than best-effort, tiered privacy,
and sovereign custody. Table~\ref{tab:gov} states honestly what is enforced today
against the target. Source attribution, structural integrity validation (CRS
enforcement, null and degenerate-geometry drops, non-finite sanitization),
propagation-stage validation, and labeled graceful degradation are shipped;
cryptographic content-hashing and on-chain node
attestation are the first governance work item on the roadmap. Today the twin serves
only L1 (environmental) data, so no L2/L3 export path yet exists to gate --- the
privacy tiering is the schema the future ISAC feed will be born into.

\begin{table*}[t]\footnotesize\centering
\caption{Zero-trust integrity gate: target versus what is enforced today.}\label{tab:gov}
\begin{tabular}{@{}p{3.2cm}p{7.3cm}p{4.9cm}@{}}
\toprule
Control & Today in code & Gap to target\\
\midrule
Source attribution & \checkmark\ \texttt{source} on every FeatureCollection plus \texttt{catalog.json}; \texttt{scene\_manifest.json} carries projected geometry, height/elevation, tower, material, and solver metadata into RF outputs & ---\\
Content hashing & pending & SHA-256 per cube at ingest, on-ledger\\
Node attestation & pending & Requires Amini Chain backbone\\
Integrity validation gate & Structural checks plus projected-metre CRS, terrain/elevation, solver-stage, and frequency/material-validity checks & No cryptographic admission gate or field-calibration gate\\
Graceful degradation & \checkmark\ gov-grid $\rightarrow$ census proxy, labeled; empty-but-valid catalog & ---\\
Sovereign custody & \checkmark\ self-hosted store and CPU-only Mitsuba/Sionna solve; no cloud, CUDA, or GPU dependency & Formalize multi-node operating policy\\
\bottomrule
\end{tabular}
\end{table*}

\subsection{Reproducibility}
\paragraph{National ingest.} The served store is fully regenerable from the
committed shapefiles by a single idempotent pipeline (\texttt{twin\_ingest}). Per
layer it reprojects to EPSG:4326, removes null or degenerate geometry, applies a
declarative rename/keep specification, writes \texttt{barbados\_twin.gpkg}, emits
PMTiles for high-count layers, and records source, CRS, count, and bounding box in
\texttt{catalog.json}. Re-running deletes and rebuilds the store rather than
performing an undocumented manual merge.

\paragraph{Propagation build.} A second one-command workflow in
\texttt{ulap-digital-twin}/\texttt{ulap-scope} executes ten out-of-process stages
across three pinned environments: GeoPandas/GDAL preparation, Blender~4.4
construction/export, and Sionna RT~2.0.1 solution. The solve uses the Mitsuba LLVM
backend on CPU; CUDA is not required. Stage contracts are carried by
\texttt{scene\_manifest.json}, and automated tests plus continuous integration
exercise the pipeline. Thus both the served geodata and the propagation scene are
rebuildable on commodity sovereign hardware. The Phase~1 work item extends both
manifests with SHA-256 hashes, ingest/build timestamps, license, and privacy tier.
The public release described in \S\ref{sec:code-availability} will package these
workflow and environment contracts with the paper's derived artifacts.

\paragraph{Claims register.} Every quantitative claim in this paper is bound to
the artefact that produces it in a machine-checkable register
(\texttt{claims/claims.yaml}), and a single command re-executes the producers
and compares each reported value against its artefact within a declared
tolerance. Claims that cannot be verified from the public release --- those
derived from non-redistributable government data, such as the island-wide layer
counts --- are recorded in the register as unverifiable with the reason stated,
rather than omitted.

\subsection{Computational cost and sovereign-hardware feasibility}
\label{sec:bench}

The sovereignty argument of \S\ref{sec:security} rests on a practical claim: that the
deterministic workflow runs on hardware a national institution can procure and operate,
without CUDA and without a commercial cloud. That claim is only meaningful if it carries
numbers, so we report them.

Every propagation stage was executed on two deliberately dissimilar machines --- an NVIDIA
GB10 (\texttt{aarch64}, 20 logical cores, 122~GB) and an NVIDIA H200 NVL (\texttt{x86\_64},
Intel Xeon 6760P, 256 logical cores, 503~GB) --- under a matched software stack
(Sionna~RT~2.0.1, Mitsuba~3.8.0, Dr.Jit~1.3.1). Each cell was run three times per backend;
we report the median with the full observed spread, never a single timing. The Mitsuba
variant selected at runtime was read back from every run rather than assumed, because
Sionna attempts the CUDA variants first and will silently execute a nominally CPU run on a
GPU if one is visible. A CPU-labelled measurement that selected a CUDA variant is treated as
invalid rather than relabelled.

\begin{table*}[t]\footnotesize\centering
\caption{Propagation-stage wall clock, seconds. Median of three runs, full spread in
parentheses, Sionna~RT~2.0.1 on both hosts. CPU rows use the Mitsuba LLVM backend with CUDA
devices hidden. Timings correspond to the solver settings then in force (coverage at 5~m
cells, $10^7$ samples; ground-following at $10^6$); the released defaults have since been
raised for figure honesty, which lengthens those stages accordingly.}
\label{tab:bench}
\begin{tabular}{@{}lrrrr@{}}
\toprule
Stage & GB10 CPU & GB10 CUDA & H200 CPU & H200 CUDA\\
\midrule
Coverage, flat      & 8.56 (8.47--8.69)  & 3.45 (3.41--3.61) & 4.48 (4.27--4.48) & 3.84 (3.81--3.87)\\
Coverage, terrain   & 9.98 (9.79--11.23) & 3.56 (3.48--3.56) & 4.25 (4.11--4.46) & 3.95 (3.90--3.96)\\
Frequency analysis  & 3.41 (3.31--3.77)  & 2.74 (2.62--2.93) & 4.92 (4.91--4.95) & 3.96 (3.94--3.96)\\
mmWave / SINR       & 2.59 (2.00--2.80)  & 2.27 (2.24--2.28) & 3.21 (3.21--3.28) & 2.88 (2.86--2.90)\\
Ground-following    & 1.93 (1.91--2.05)  & 1.78 (1.70--1.83) & 2.26 (2.25--2.39) & 2.02 (2.01--2.04)\\
\midrule
\textbf{Full pipeline} & \textbf{26.5} & \textbf{13.8} & \textbf{19.1} & \textbf{16.6}\\
\bottomrule
\end{tabular}
\end{table*}

Three observations follow.

First, the complete study finishes in tens of seconds and peak resident memory never exceeds
$1.25$~GB on any host or backend. The binding constraint on reproducing this work is
therefore neither compute nor memory but access to the authoritative geodata.

Second, the GPU advantage depends strongly on sampling fidelity. At the settings of
Table~\ref{tab:bench} it is modest --- roughly $2.5\times$ on the coverage solves and close
to unity on the lighter stages, which are dominated by process start-up and just-in-time
compilation rather than ray throughput --- and it is this modest gap at survey settings that
makes the CPU-only claim practical rather than merely technically true. At the raised
sampling default the picture inverts (see below): the ray-bound solve dominates and the gap
widens to ${\sim}36\times$. Neither figure is a GPU throughput measurement and neither
should be read as one. The claim also holds
on \texttt{aarch64}, an architecture the original workflow did not target. The division of
labour sharpens at high sampling fidelity: the 57-plane ground-following stack at $10^8$
samples per transmitter solves in $7.3$~s on the GPU against $262$~s on the CPU backend
($\sim$36$\times$). CPU-only execution is what makes the pipeline sovereign; the GPU is what
makes honest sampling cheap.

Third, a portability caveat that cost us a full diagnostic cycle and is worth stating so
others do not repeat it. On the H200 host every CPU cell initially failed with
\texttt{ImportError: \ldots the LLVM backend is inactive because the LLVM shared library
("libLLVM.so") could not be found}. This is not an architectural limit: the distribution
ships \texttt{/lib64/libLLVM.so.20.1} but no unversioned \texttt{libLLVM.so}, which is the
name Dr.Jit attempts to load. Setting \texttt{DRJIT\_LIBLLVM\_PATH} to the versioned library
makes all CPU cells pass. The CPU-only execution path therefore carries an undeclared
dependency on LLVM being discoverable, and a deployment that assumes CPU portability without
verifying it may silently acquire a GPU requirement.

\paragraph{Determinism.} The pipeline's only numeric artefact,
\texttt{link\_metrics.csv}, is byte-identical across both architectures, both backends
\emph{and} both solver versions tested. Because that file is written at two decimal places,
this establishes agreement to $0.01$~dB and $0.01$~ns with identical path counts --- not
bit-identical floating point, a distinction we make explicitly.
Concretely, the same artefact is byte-identical on an NVIDIA GB10
(\texttt{aarch64}), an NVIDIA H200~NVL (\texttt{x86\_64}) and a Raspberry~Pi~4
(\S\ref{sec:commodity}): MD5 \texttt{08afea6d\ldots} on all three and on the
archived copy. Rendered figures are not
byte-stable: PNG outputs differ across hosts and backends in $0.2$--$2.4$\,\% of pixels with
mean absolute difference below $0.04/255$, concentrated at colormap boundaries and glyph
edges. Reproduction of the numeric result is therefore verifiable by checksum; reproduction
of the figures is not, and should not be claimed.

\subsection{Comparison against baseline propagation models}
\label{sec:comparison}

\S\ref{sec:rt-novelty} argues that deterministic ray tracing is an upgrade on stochastic
national planning. That is a comparative claim, and the preceding sections do not test it:
the results in Table~\ref{tab:newton-results} are reported against no external baseline. We
therefore evaluate the solver against a closed-form reference on identical geometry, and
quantify the internal ablation the earlier sections describe qualitatively, and
test the stochastic-planning claim directly. Metrics were
fixed before any comparison was executed.

\paragraph{Against a closed-form model.} On the Rising~Sun transect a free-space plus
two-ray model with knife-edge diffraction, evaluated on the same manifest, same tower and
same fourteen receiver positions, tracks the ray tracer to \textbf{0.11~dB RMS} with a mean
signed bias of $-0.05$~dB, a fitted path-loss exponent of $1.739$ against the ray tracer's
$1.732$ ($\Delta n = 0.007$), and a Pearson correlation of $0.99991$.

We report this agreement plainly rather than concealing it, and we also report why it
occurs, because the number is misleading on its own. That radial is rural line-of-sight; the
median resolved path count is \emph{two} --- direct plus ground reflection --- and the
pipeline forms path gain as an incoherent sum of per-path powers. Under those conditions the
deterministic solve reduces to the phase-averaged two-ray model by construction. The
knife-edge diffraction term contributed exactly $0$~dB at all fourteen points, so the
analytical model's only building-aware mechanism never engaged. This is an easy case for the
closed-form model, and it should be read as bounding where determinism is \emph{not}
required rather than as evidence against it.

The material difference is multipath structure, which the closed-form model cannot represent
at all. Where a third, building-reflected path appears, the ray tracer resolves an RMS delay
spread of $361.1$~ns at $450$~m and $196.9$~ns at $950$~m against a median of $0.31$~ns
elsewhere; the analytical model returns no delay-spread estimate at any range. A planner
using the closed-form model alone would site a link that is acceptable on path gain and
exposed to inter-symbol interference at precisely those positions.

\paragraph{The terrain ablation, quantified.} Table~\ref{tab:ablation} replaces the
qualitative claim that relief ``isolates the effect'' with magnitudes, computed on one
common $8$~m grid with three seeds per variant.

\begin{table}[t]\footnotesize\centering
\caption{Coverage statistics by scene variant, best-server path gain in dB, on one common
8~m grid at $10^8$ samples per transmitter --- the raised default at which sample starvation
no longer masquerades as shadow. Ground-following uses the corrected ceiling-plane construction ($K=57$, receiver never below local ground); the superseded nearest-plane stack reported $28.1$\,\% no-coverage, most of it discretisation.}
\label{tab:ablation}
\begin{tabular}{@{}lrrrr@{}}
\toprule
Variant & Median & $p_{10}$ & $p_{90}$ & No coverage\\
\midrule
Flat              & $-113.56$ & $-127.75$ & $-93.61$ & 0.38\,\%\\
Terrain plane     & $-112.30$ & $-125.71$ & $-94.16$ & 0.13\,\%\\
Ground-following  & $-116.51$ & $-128.14$ & $-93.86$ & 1.16\,\%\\
\bottomrule
\end{tabular}
\end{table}

The medians barely move --- about $-0.2$~dB between adjacent variants --- but that summary
conceals the effect. $16.3$\,\% of cells differ by more than $10$~dB, $p_{10}$
moves by $-12.6$~dB, and the \emph{serving cell} changes in $10.0$--$22.3$\,\% of cells
depending on the pair compared, against a Monte-Carlo noise floor of $0.99$\,\% established
by re-solving identical configurations --- a $22.5\times$ signal. We note that these figures
are materially smaller than the same statistics computed at the solver's former sampling
default: at $10^6$ samples per transmitter, more than half of the apparent serving-cell
churn was estimator noise rather than terrain. Raising the default was what revealed it. Terrain therefore matters materially for association
and cell-edge behaviour while being nearly invisible in an area median --- which is an
argument for reporting association statistics rather than medians in coverage studies.

\paragraph{Correcting the ground-following construction.} Re-examining this map found
that most of its dark area was measurement error rather than shadow, and we report the
correction rather than the original figure.

The construction stacked nine horizontal planes across $55.5$~m of relief and selected, per
cell, the plane \emph{nearest} \mbox{terrain~$+\,1.5$~m}. Nine planes over that relief is
$6.94$~m spacing, so the selected plane sat a median $1.81$~m from the intended height and,
in $28.97$\,\% of cells, \emph{below} local ground --- where the receiver is occluded by the
terrain itself and returns no coverage. Selecting the nearest plane above the target instead
of the nearest plane in either direction removes this entirely, and deriving the plane count
from a stated height tolerance ($K=\lceil\text{relief}/1\,\text{m}\rceil+1=57$) bounds the
residual offset to under a metre. Underground cells fall from $28.97$\,\% to zero; with the sampling default also raised to
$10^8$ per transmitter, the no-coverage fraction falls from $28.12$\,\% to $1.16$\,\% ---
the remainder of the old dark area was sample starvation, not shadow.

A second effect accounts for most of the remainder. A sweep of solver samples per
transmitter shows the no-coverage fraction falling roughly tenfold per decade with no floor
until $10^8$--$10^9$: at the $10^6$ this stage shipped with, $47$\,\% of the map is
\emph{unsampled} rather than shadowed. Only at $10^9$ does the residual resolve into
geometry.

With both effects removed, genuine ridge shadowing is small. An independent pure-geometry
line-of-sight test against the DTM --- no ray tracer involved --- puts it at $0.75$\,\% of
the map, and the ray tracer resolves less still because diffraction fills most of it in.
The effect is therefore real and correctly attributed to relief, but roughly two orders of
magnitude smaller than the uncorrected map suggests.

We draw the honest conclusion: the quantitative case for the ground-following surface is
\emph{not} shadow area. It is that accounting for relief changes the serving cell in
$22.3$\,\% of cells against a $0.99$\,\% Monte-Carlo noise floor --- a $22.5\times$ signal ---
and moves $16.3$\,\% of cells by more than $10$~dB. That is a planning-relevant result;
the shadow area was largely an artefact of how we measured.

\begin{figure}[t]\centering
\includegraphics[width=\linewidth]{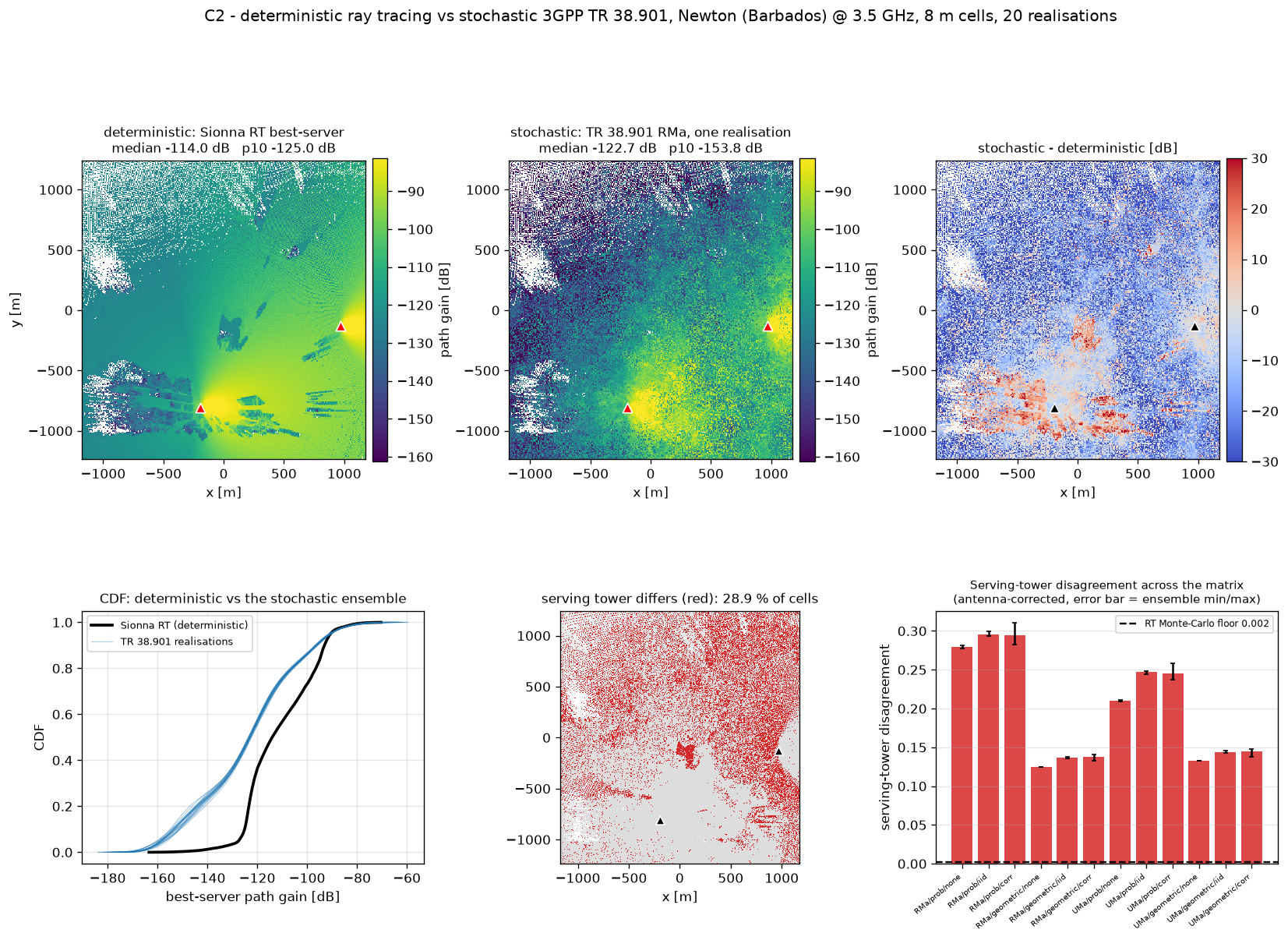}
\caption{Deterministic best-server path gain against the TR~38.901 RMa
stochastic planning surface on identical geometry ($83{,}282$ cells, 20
shadow-fading realisations). Evidence: \texttt{comparison/results/c2.json}.}
\label{fig:c2-stochastic}
\end{figure}

\paragraph{Against a stochastic planning surface.} We implemented 3GPP TR~38.901
RMa and UMa from the standard --- every constant as printed, nothing fitted ---
on the identical grid, towers and frequency, with spatially correlated shadow
fading over 20 seeded realisations; the deterministic reference is the
best-server map, both surfaces carry the same antenna correction, and
$83{,}282$ cells are compared per realisation. On the primary RMa cell the two
surfaces measurably differ: RMS error $19.99$~dB (ensemble median;
$18.94$--$21.37$ across realisations), mean signed bias $-12.69$~dB (TR~38.901
pessimistic), Pearson $r$ of $0.648$, and $55.8\,\%$ of cells more than $10$~dB
apart. The disagreement concentrates where planning decisions are made:
$29.4\,\%$ of cells select a \emph{different serving mast} --- $135\times$ the
ray tracer's own $0.217\,\%$ Monte-Carlo noise floor --- and the cell-edge
$p_{10}$ levels sit $28$~dB apart ($-125.0$ against $-153.1$~dB).
Fig.~\ref{fig:c2-stochastic} maps the disagreement; the full design matrix is
in \texttt{comparison/results/c2.json}.

Roughly half of the disagreement is one term. Substituting the ray tracer's own
LOS oracle --- a visibility-only solve --- for the standard's distance-based LOS
probability drops the RMS from $20.0$ to $11.0$~dB and the serving-mast
disagreement from $29.4\,\%$ to $13.7\,\%$: the stochastic surface's largest
single defect on this scene is not its propagation mathematics but that it does
not know where the buildings are. That diagnostic deliberately violates the
standard's stochastic design and is not quotable as TR~38.901 performance.

The verdict is deliberately narrow. These numbers establish that the two
methods \emph{differ} far beyond solver noise, and where the difference lands;
they do \emph{not} establish that the deterministic map is more accurate.
Nothing in this study is calibrated against field measurement, and of the two
models, TR~38.901 is the one with an empirical pedigree. What determinism
demonstrably changes is serving-cell association and cell-edge behaviour;
whether it improves them awaits calibration, which remains future work.

\subsection{How small can the machine be?}
\label{sec:commodity}

Sovereignty that presupposes a data-centre GPU fleet is sovereignty contingent on export
licences, foreign currency and a vendor relationship. The useful question is therefore not
how fast the pipeline runs, but how small a machine still runs it. We answer it on the
smallest machine we could obtain rather than by extrapolation.

The complete deterministic workflow was executed on a \textbf{Raspberry Pi~4 Model~B}
(4$\times$Cortex-A72, \texttt{aarch64}, 8~GB, Debian~13, Python~3.13, microSD storage),
CPU-only with CUDA devices hidden and the Mitsuba LLVM backend forced. The entire solver
stack --- Sionna~RT~2.0.1, Mitsuba~3.8.0, Dr.Jit~1.3.1 --- installed from stock PyPI wheels
with nothing compiled from source.

\begin{table}[t]\footnotesize\centering
\caption{Full pipeline on a Raspberry Pi~4 Model~B, CPU-only, no GPU.}
\label{tab:commodity}
\begin{tabular}{@{}lrrr@{}}
\toprule
Stage & Wall (s) & Peak RSS (MB) & CPU\\
\midrule
Coverage, flat        & 118.6 & 810 & 361\,\%\\
Coverage, terrain     & 142.1 & 814 & 367\,\%\\
Frequency analysis    &  22.2 & 505 & 180\,\%\\
mmWave / SINR         &  13.3 & 308 & 160\,\%\\
Ground-following      &  14.2 & 281 & 250\,\%\\
\midrule
\textbf{Full pipeline} & \textbf{310.4} & \textbf{814} & \\
\bottomrule
\end{tabular}
\end{table}

The complete study finishes in a little over five minutes. More consequentially, the
resulting \texttt{link\_metrics.csv} is \emph{byte-identical} to the file produced on the
NVIDIA GB10 and the NVIDIA H200~NVL, and to the copy archived with the code release: MD5
\texttt{08afea6d\ldots} on all four. A single-board computer costing on the order of one
hundred US dollars therefore reproduces the numeric result of this paper exactly, to the
0.01~dB at which that file is written.

Peak resident memory was 814~MB on the Pi, essentially unchanged from every other machine
tested: the workflow's footprint is fixed by the scene, not by the host. Corroborating
runs on \texttt{aarch64} under hard \texttt{cgroup} limits completed at four cores with a
4~GB ceiling and again at two cores with 2~GB, with memory invariant throughout.

\paragraph{What this costs.} The Pi is approximately $12\times$ slower than the 20-core
server-class ARM host on the same CPU-only work (310~s against 26.5~s for the full
pipeline). That is the honest price of the sovereignty argument: the capability is
\emph{affordable}, not fast. We also note that the board reached 68--73~$^\circ$C and
\texttt{vcgencmd} reported that the soft temperature limit had been reached, so these
timings should be read as achievable on a passively-cooled board rather than as a best
case, and that each stage was run once rather than as a median of three.

\subsection{Towards-6G deterministic propagation methodology}
\label{sec:rt-method}
The deterministic workflow groups its ten executable stages into four
methodological phases:
\begin{enumerate}
\item \textbf{Clip.} \texttt{clip\_study\_area.py} extracts a 2~km box centered on
The Barbados Heritage District at Newton Plantation ($-59.5339^\circ$, $13.0881^\circ$) while preserving
EPSG:21292. Towers from the August~2023 national antenna register are captured
with an 800~m margin.
\item \textbf{Preprocess.} \texttt{preprocess\_scene.py} emits
\texttt{scene\_manifest.json}: 576 footprints with per-building LiDAR-derived
height and ground elevation; a 70$\times$70 terrain grid interpolated from the
building DTMs (54.7--110.3~m, approximately 55~m relief); an ESRI satellite
basemap warped from EPSG:3857 to EPSG:21292; and mast positions seated on sampled
ground elevation.
\item \textbf{Build/export.} \texttt{build\_scene.py} extrudes footprints to
their measured heights and seats them on terrain (or $z=0$ for the controlled
flat variant). \texttt{export\_mitsuba.py} writes Z-up Mitsuba XML and explicitly
tags buildings as \texttt{itu\_concrete} and terrain as
\texttt{itu\_medium\_dry\_ground}.
\item \textbf{Solve.} Sionna RT evaluates Newton (30~m monopole on 69~m ground)
and Rising Sun (24~m monopole on 100~m ground) in-scene, with Boarded Hall and
Oistins represented as external interferers where required. The same manifest
drives flat, terrain, frequency, multi-tower, transect, and moving-receiver
experiments.
\end{enumerate}

This construction is the deterministic upgrade of the stochastic island-wide
planning surface in Figure~\ref{fig:rf-coverage}. Standard map-derived workflows
often begin with generic OSM geometry; here, footprints, LiDAR heights, ground
elevations, terrain, and the tower inventory remain traceable to sovereign
government data through the propagation solve.

\subsection{Newton propagation results}
\label{sec:rt-results}
Table~\ref{tab:newton-results} reports the common-scene experiments; the maps in
Figures~\ref{fig:freq-sweep}--\ref{fig:multitower} show representative outputs.
All values are deterministic simulation results rather than field measurements.

\begin{table*}[t]\scriptsize\centering
\caption{Newton towards-6G propagation studies on the authoritative scene.
Path-gain and delay values are simulated and not yet field-calibrated. Solver settings are
stated \emph{per row}: they differ between studies, and quoting one row's configuration
against another's numbers will not reproduce them.}
\label{tab:newton-results}
\begin{tabular}{@{}p{2.8cm}p{4.7cm}p{8.2cm}@{}}
\toprule
Study & Setup & Result\\
\midrule
Best-server coverage & 3.5~GHz; TR~38.901-style transmit pattern; depth 5;
8~m cells; $10^8$ samples/transmitter & Building shadowing is resolved in the
dense settlement southwest of Newton; controlled flat and terrain-draped
variants isolate the effect of relief.\\
Frequency sweep & 1.8/3.5/6/10~GHz; Newton and Rising Sun; depth 5;
8~m cells; $10^6$ samples/transmitter & Median best-server
path gain is $-106.2/-112.6/-117.2/-121.5$~dB, a 15.3~dB penalty from 1.8 to 10~GHz on
identical geometry. The sweep stops at 10~GHz because
\texttt{itu\_medium\_dry\_ground} is defined only over 1--10~GHz.\\
mmWave & 28/60~GHz; scene explicitly retagged concrete-only & Median path gain is
$-130/-136$~dB and useful coverage contracts to line-of-sight lobes around the
masts, directly illustrating the densification pressure at FR2.\\
Multi-tower association & Four towers; 33~dBm/sector; 100~MHz;
interference-limited & Best-server association exposes a cell-edge interference
zone west of Newton and SINR pockets above 25~dB near Newton and Rising Sun.\\
Link transect & Rising Sun to scene center; 50--1350~m; receiver at 1.5~m & Path
gain declines from $-98.9$ to $-123.8$~dB. Most locations have two paths (LoS and
ground); a third building-reflected path produces 361~ns RMS delay spread at
450~m and 197~ns at 950~m.\\
Ground-following map & Nine receiver planes span the 55~m relief; per cell choose
the plane nearest terrain$+1.5$~m & Produces a true 1.5~m-above-ground-level map
and reveals ridge shadowing absent from a single horizontal receiver plane.\\
Drive test & 60-step road route; fresh path solve at each step & The
\texttt{moving\_rx.gif} artifact records live best-server gain and handover
events along the route.\\
\bottomrule
\end{tabular}
\end{table*}

\subsection{Methodological novelty and limits}
\label{sec:rt-novelty}
\paragraph{A provenance-preserving physics ladder.} The RF scene twin now spans a flat approximation, a stochastic
national-planning surface, and deterministic site studies under one sovereign
source contract. Measured head-to-head (\S\ref{sec:comparison}), the
deterministic surface changes serving-cell association and cell-edge behaviour
materially --- a different serving mast in $29.4\,\%$ of cells and cell-edge
$p_{10}$ levels $28$~dB apart --- but whether it \emph{improves} on the
stochastic surface is not established: neither map is calibrated against field
measurement, and calibration remains future work. Consequently, a
coverage pixel can be traced to its EPSG:21292 coordinate, government feature,
height/elevation attributes, material assumption, solver configuration, and
output run. This gives layer~4 of the S-CDT a physics engine and provides the
concrete $\mathbf{H}_{\mathrm{background}}$ surrogate that a later EKF/ISAC loop
can calibrate.

\paragraph{Terrain and material validity.} Sionna's radio-map receiver surface is
horizontal; over 55~m of local relief, that does not maintain a constant
above-ground receiver height. \texttt{sionna\_terrain\_ground.py} works around
this limitation using a stack of horizontal receiver planes whose count derives from a
stated height tolerance ($K=\lceil\text{relief}/1\,\text{m}\rceil+1=57$ here), selecting
per cell the nearest plane \emph{at or above} the target height so the receiver can never
sit below local ground.
Frequency claims are also bounded by material validity: mixed
concrete/medium-dry-ground studies stop at 10~GHz, while 28/60~GHz outputs are
explicitly labeled concrete-only rather than silently extrapolating the ground
model.

\paragraph{Sovereign operations and honesty boundary.} CPU-only execution makes
the pipeline reproducible on commodity national infrastructure, and the offline
scene manifest now also drives SCOPE's live UE-Sim ray-traced coverage snapshots
(implementation commits \texttt{f9e8fd31} and \texttt{f7b2a798}). The present
claim is nevertheless \emph{towards-6G propagation}: 10~GHz exercises an
upper-midband/FR3 candidate regime and 28/60~GHz the FR2 extreme, but the antennas
remain $1\!\times\!1$ V-polarized and no result is field-calibrated. What is
built is the site-specific background-channel substrate, not yet the active ISAC
sensing or cognitively closed loop.

\subsection{Deployed system}
\label{sec:deployed-system}
\begin{figure*}[p]
\centering
\captionsetup{font=scriptsize,labelfont=bf,skip=3pt}
\begin{minipage}[t]{0.485\textwidth}
\centering
\includegraphics[width=\linewidth]{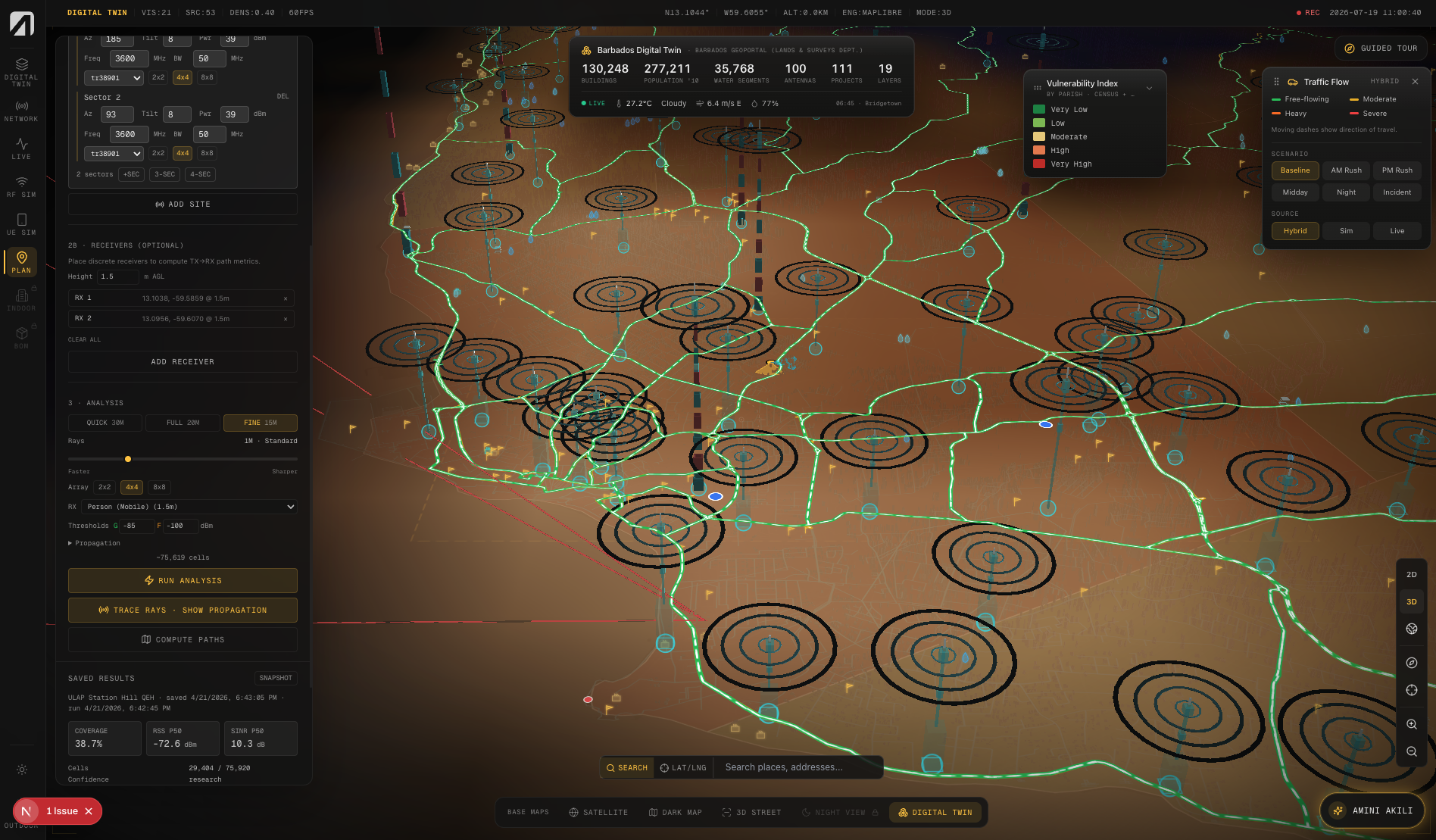}
\captionof{figure}{Network coverage planning in the twin's PLAN mode. Per-sector
RF configuration (azimuth, tilt, and power; 3600~MHz, 50~MHz bandwidth) is
propagated with a 3GPP TR~38.901 channel model over national infrastructure and
the parish-vulnerability surface. Discrete receivers yield per-run statistics:
38.7\% coverage, $-72.6$~dBm median RSS, and 10.3~dB median SINR across
$\sim$75{,}619 cells. Concentric rings are predicted per-site coverage cells---
the dual-function radio infrastructure repurposed for ISAC sensing in
\S\ref{sec:physical}.}
\label{fig:rf-coverage}
\end{minipage}\hfill
\begin{minipage}[t]{0.485\textwidth}
\centering
\includegraphics[width=\linewidth]{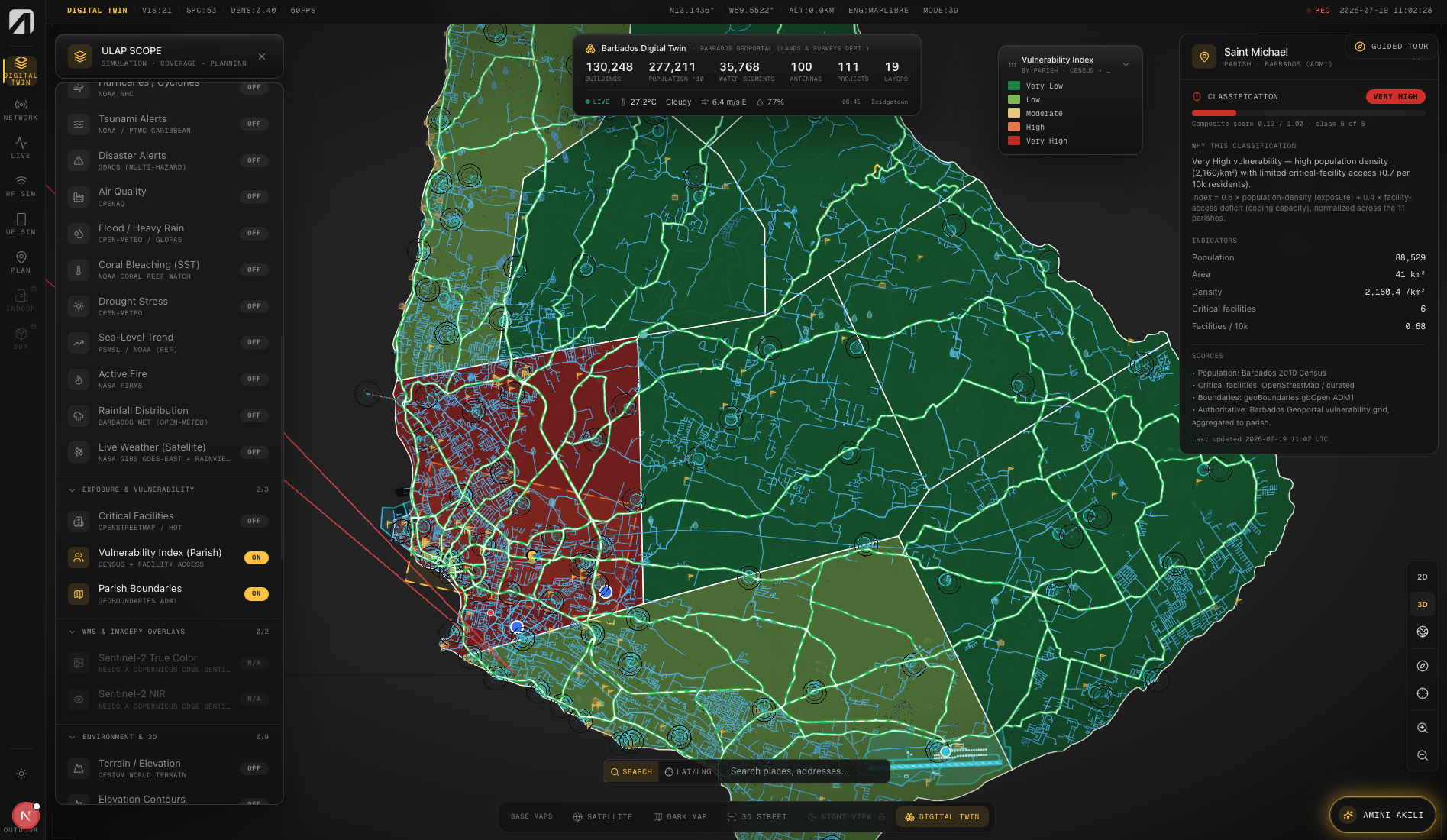}
\captionof{figure}{Authoritative per-parish vulnerability index. Parishes are
graded Very Low to Very High from the Barbados Geoportal vulnerability grid
aggregated to ADM1 boundaries. The inspected parish, Saint Michael, is Very High
(composite 0.19/1.00; class 5 of 5), reflecting population density
(2,160/km$^2$; population 88,529) and a critical-facility access deficit
(0.68 facilities per 10,000 residents). Its panel names the Barbados 2010
Census, OSM/curated critical facilities, geoBoundaries ADM1, and authoritative
Geoportal grid, realizing the labeled-provenance contract of
\S\ref{sec:impl-governance}.}
\label{fig:vulnerability-index}
\end{minipage}

\vspace{0.8em}
\begin{minipage}[t]{0.485\textwidth}
\centering
\includegraphics[width=\linewidth]{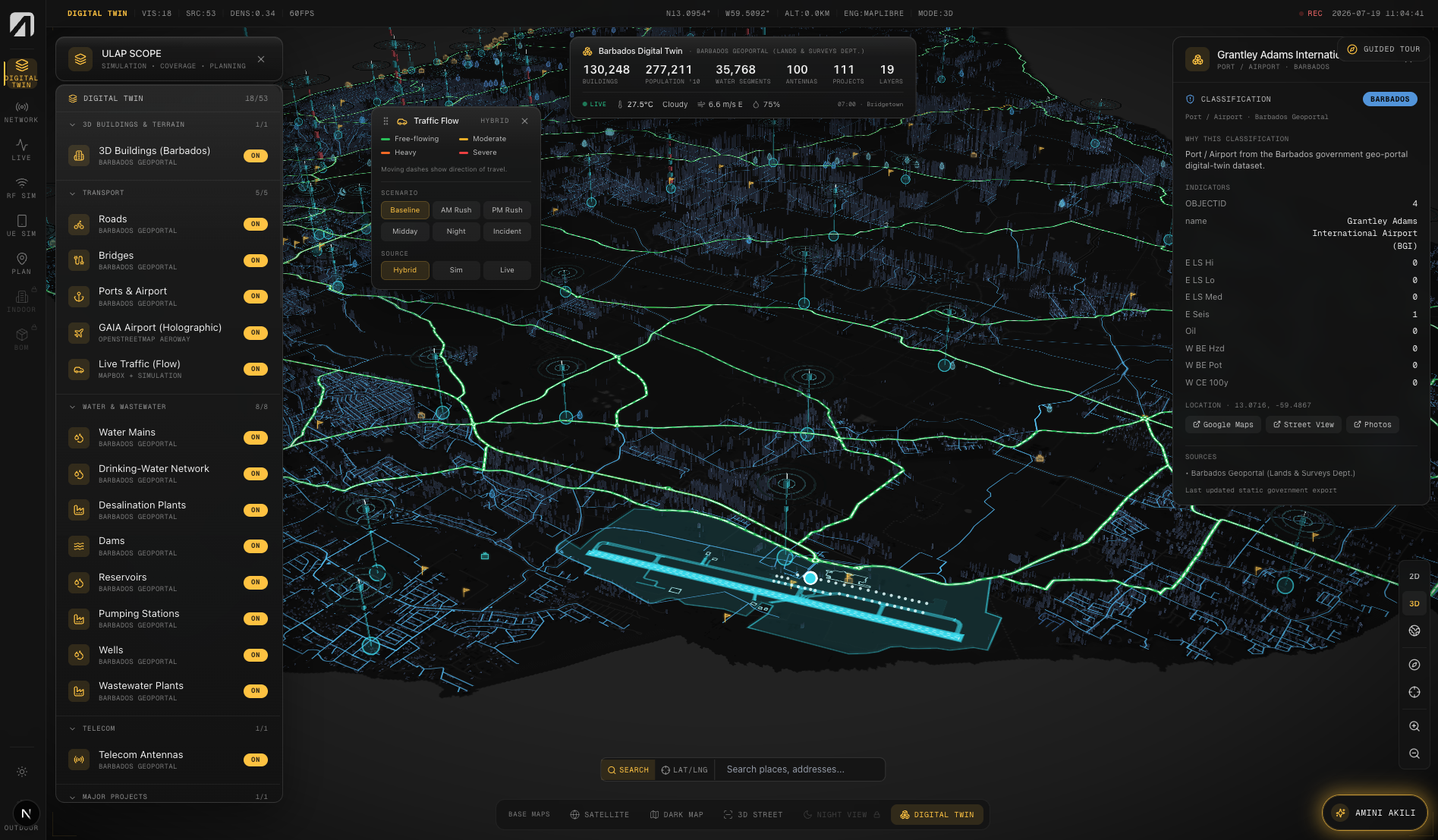}
\captionof{figure}{Government-asset inspection with hazard attributes. The
holographic 3D render selects Grantley Adams International Airport (BGI) on the
dark-map building and terrain base. Its inspect panel exposes provenance
(Barbados Geoportal) and per-asset hazard return-period fields---landslide
\texttt{E\_LS}, seismic \texttt{E\_Seis}, and coastal/rainfall flood
\texttt{W\_CE}/\texttt{W\_BE}---used to join critical assets to climate-risk
models (\S\ref{sec:impl-provenance}).}
\label{fig:airport-inspect}
\end{minipage}\hfill
\begin{minipage}[t]{0.485\textwidth}
\centering
\includegraphics[width=\linewidth]{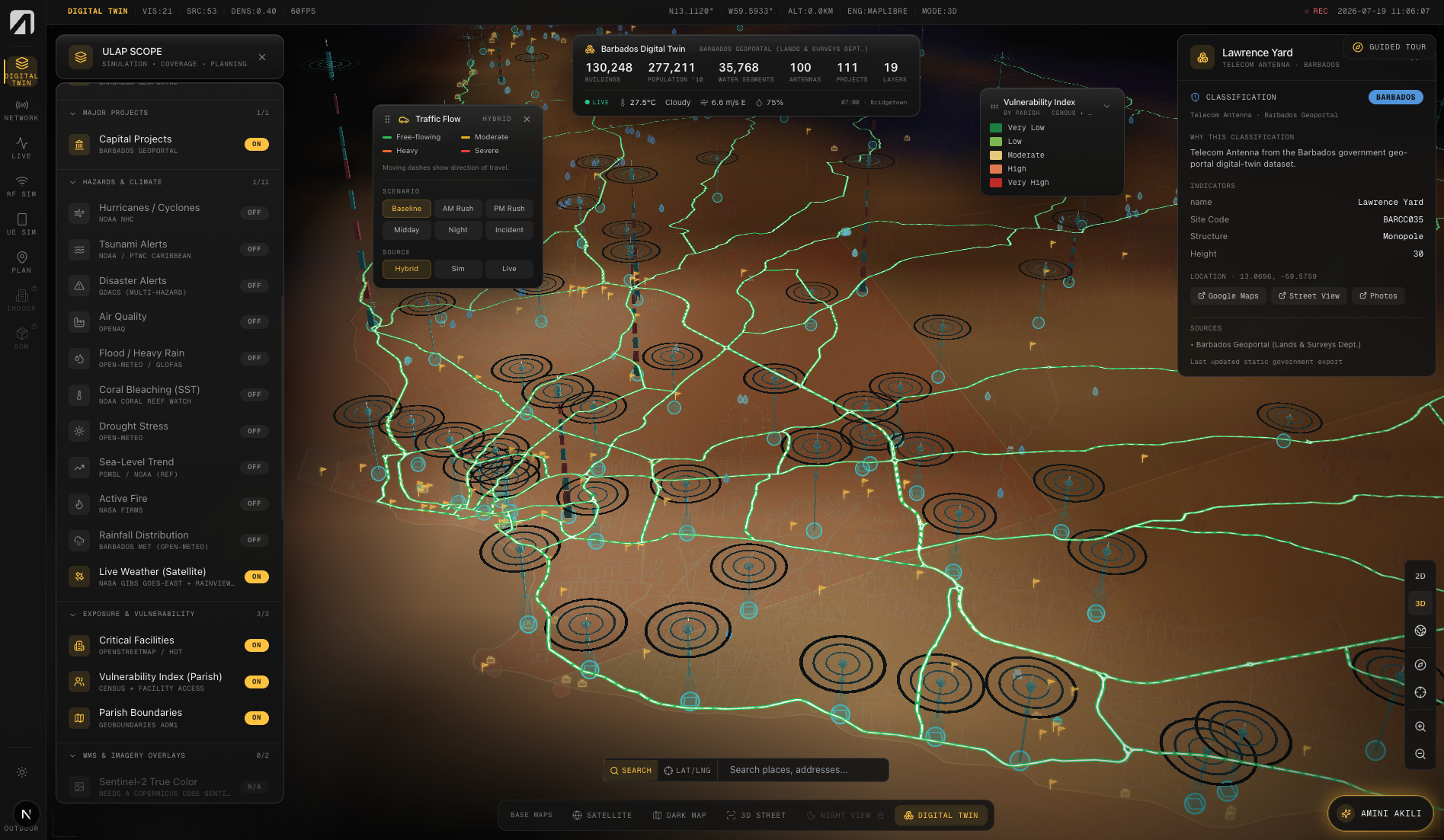}
\captionof{figure}{Telecom asset in its coverage-and-exposure context. A 30~m
monopole at Lawrence Yard (site \texttt{BARCC035}) is inspected over predicted RF
coverage cells and the parish-vulnerability surface, together with the live
weather/KPI header and traffic-flow scenario controls. This is the deployed
telecom layer that becomes the network-as-sensor substrate in the S-CDT.}
\label{fig:antenna-inspect}
\end{minipage}

\vspace{0.8em}
\begin{minipage}[t]{0.485\textwidth}
\centering
\includegraphics[width=\linewidth]{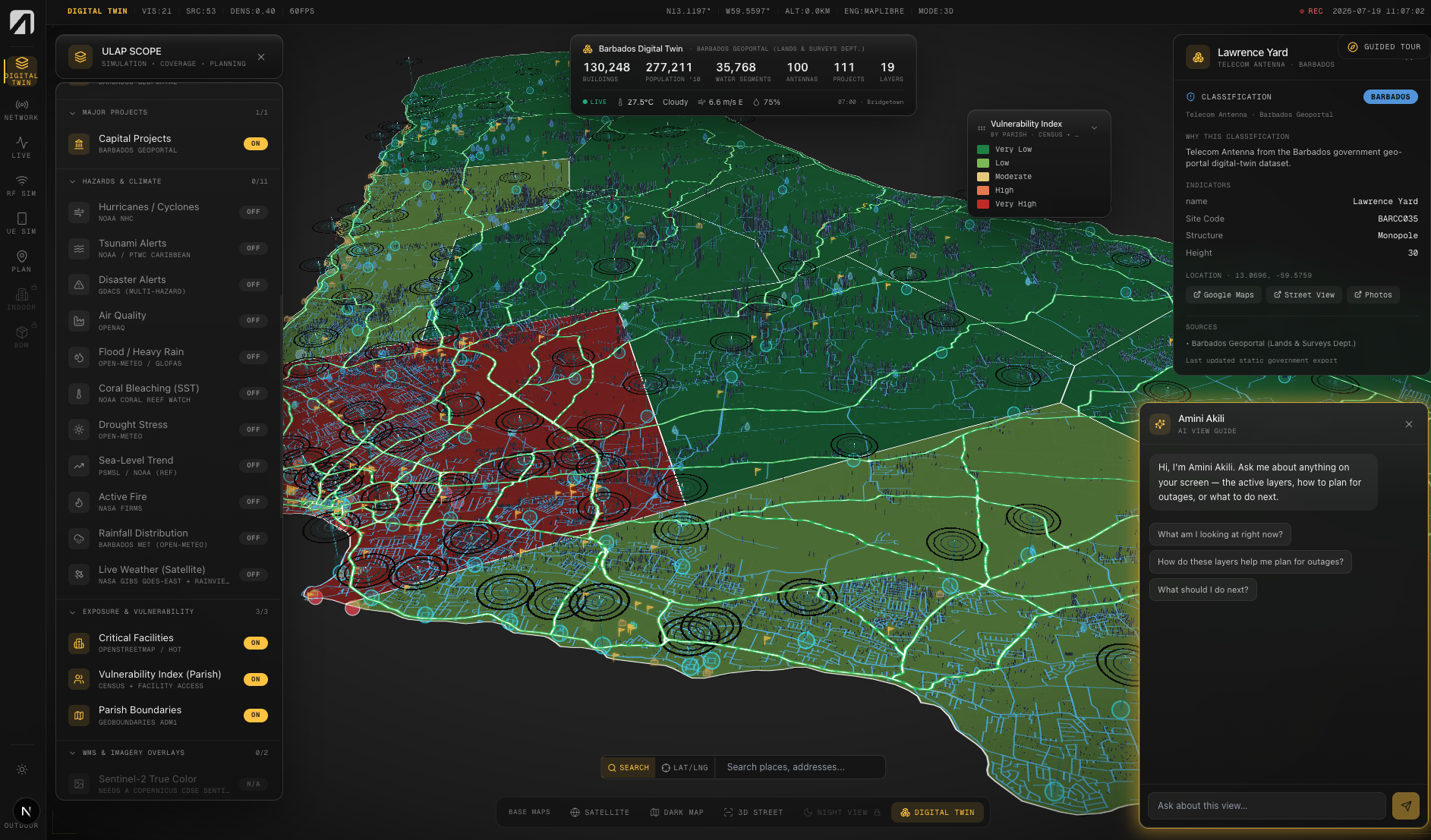}
\captionof{figure}{Natural-language interaction layer (Amini Akili). The in-map
generative assistant answers grounded questions about the current view,
including active layers, outage planning, and recommended next actions. Shown
over the parish-vulnerability choropleth with the Very High parish highlighted,
it realizes interaction layer~6 of Table~\ref{tab:map} and prefigures the GovChat
query surface in \S\ref{sec:roadmap}.}
\label{fig:akili-assistant}
\end{minipage}\hfill
\begin{minipage}[t]{0.485\textwidth}
\centering
\includegraphics[width=\linewidth]{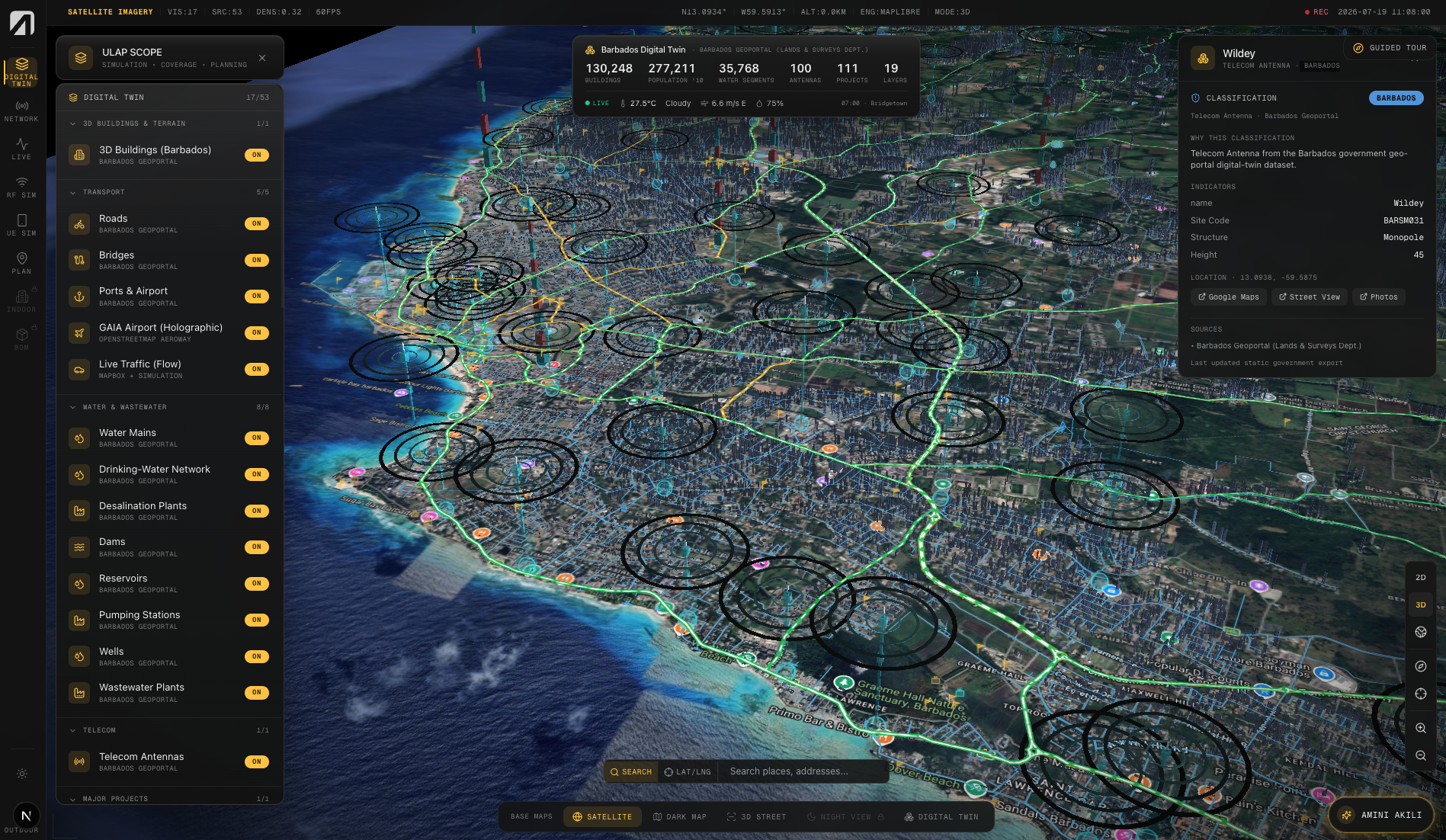}
\captionof{figure}{Twin over satellite imagery in oblique 3D. The Bridgetown and
south-coast conurbation combines the potable-water network (cyan), road network
(green), points of interest, and predicted antenna coverage cells on a satellite
basemap. Inspection of the 45~m Wildey monopole (site \texttt{BARSM031})
demonstrates multi-source fusion layered over the authoritative government
export.}
\label{fig:bridgetown-3d}
\end{minipage}
\end{figure*}

\begin{figure}[t]
\centering
\captionsetup{font=scriptsize,labelfont=bf,skip=3pt}
\begin{minipage}[t]{\linewidth}
\centering
\includegraphics[width=\linewidth]{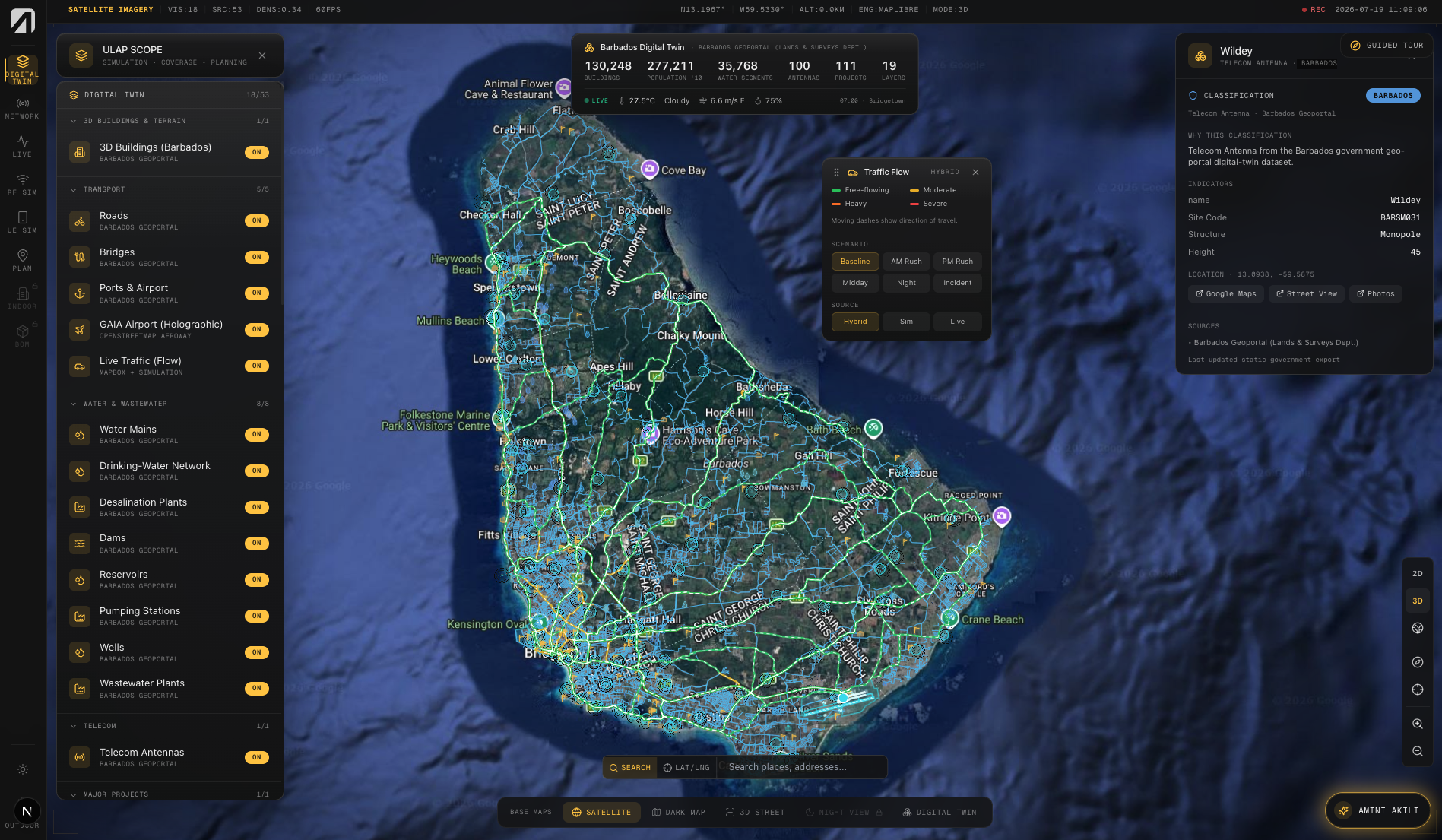}
\captionof{figure}{National context view. The full island of Barbados
(166~km$^2$) appears on a satellite basemap with complete road and water
networks, parish labels, traffic-flow scenario controls, and asset inspection.
This is the bounded, high-fidelity reference deployment generalized to the
Philippine archipelago in \S\ref{sec:deploy}.}
\label{fig:island-overview}
\end{minipage}
\end{figure}

\begin{figure*}[tp]
\centering
\captionsetup{font=scriptsize,labelfont=bf,skip=3pt}
\begin{minipage}[t]{0.35\textwidth}
\centering
\includegraphics[width=\linewidth]{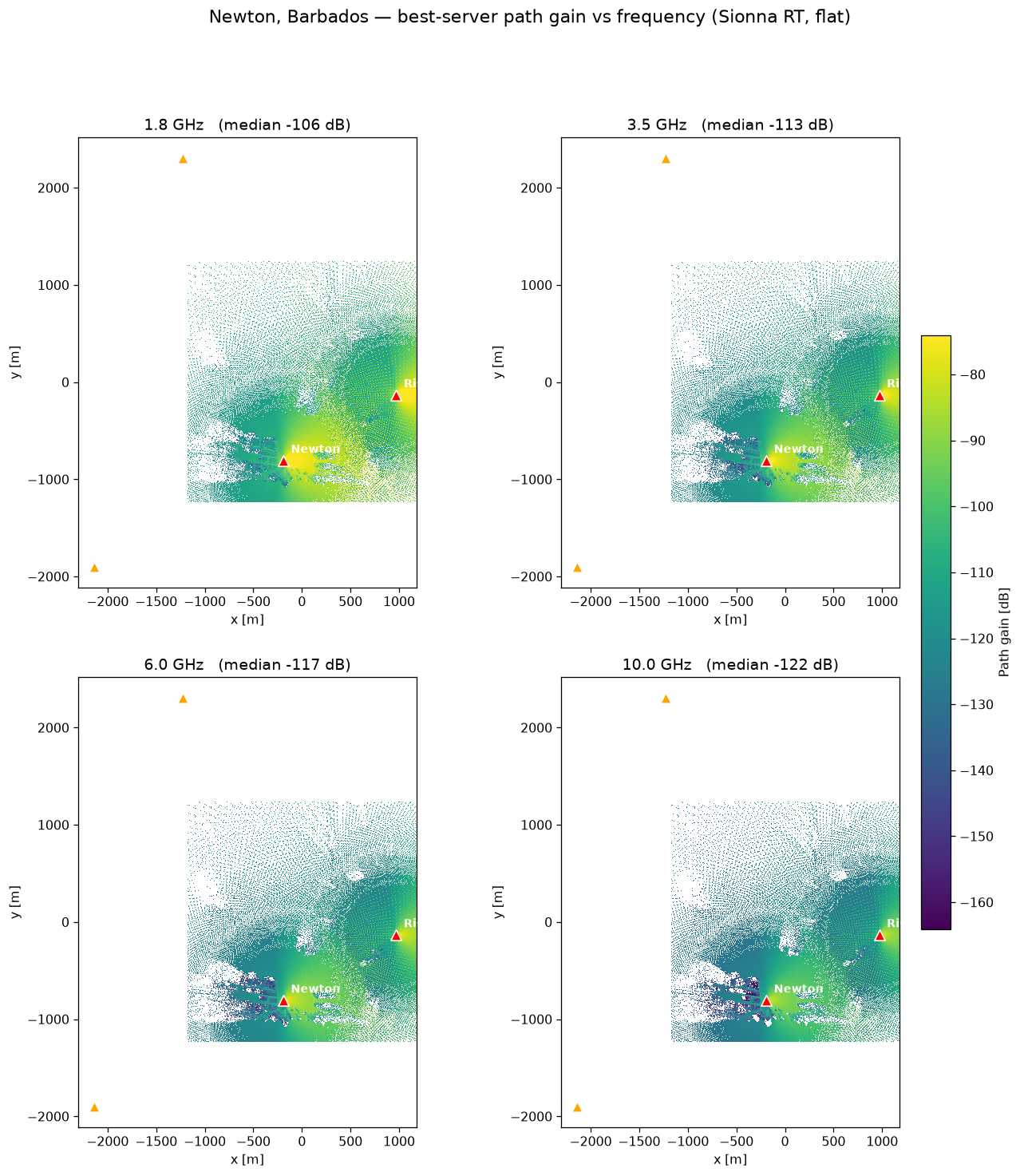}
\captionof{figure}{Frequency sweep on identical Newton/Rising Sun geometry.
Median best-server path gain falls from $-106.2$~dB at 1.8~GHz to $-121.5$~dB at
10~GHz. The mixed ground/concrete study is deliberately capped at 10~GHz, the
validity limit of \texttt{itu\_medium\_dry\_ground}.}
\label{fig:freq-sweep}
\end{minipage}\hfill
\begin{minipage}[t]{0.61\textwidth}
\centering
\includegraphics[width=\linewidth]{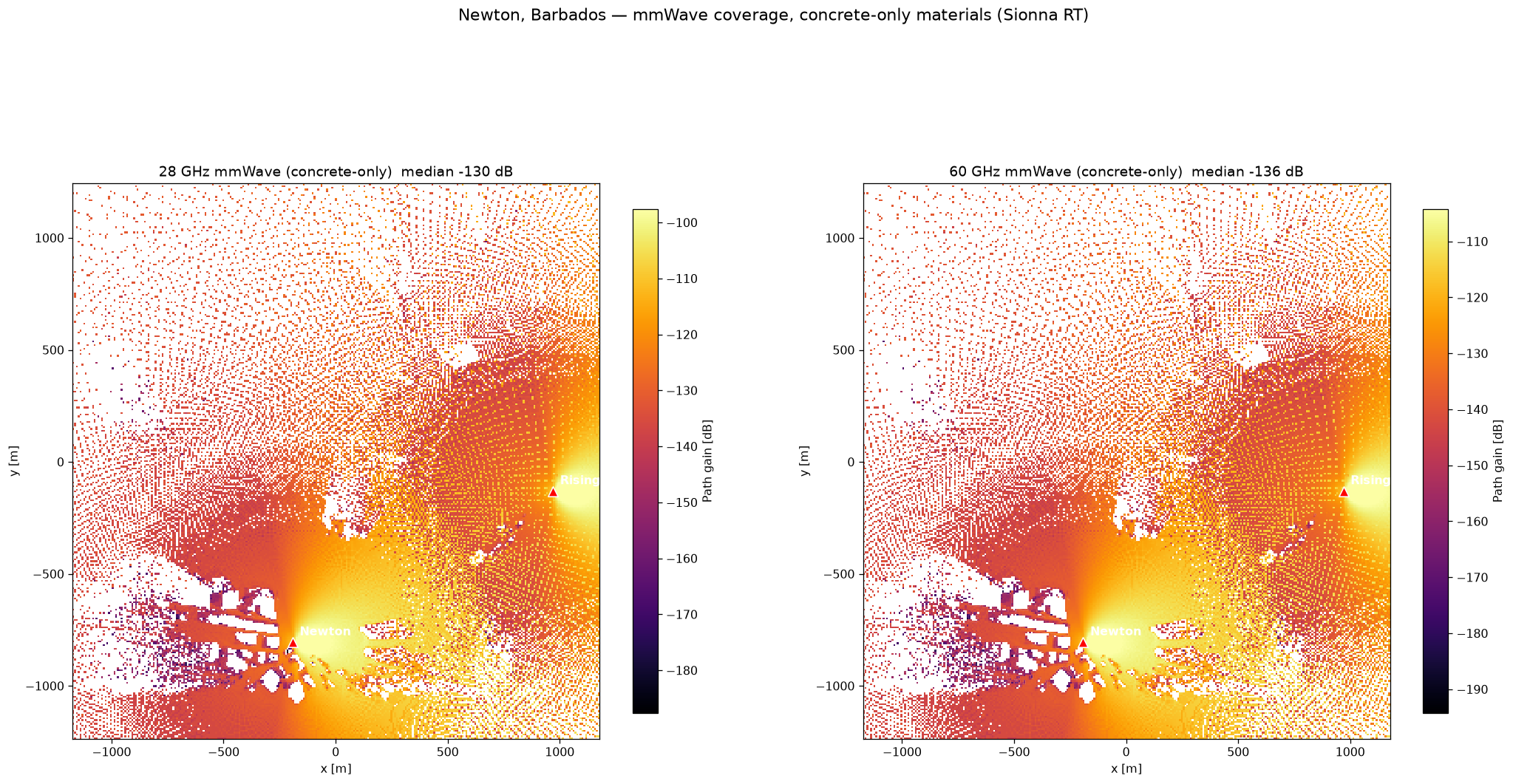}
\captionof{figure}{Concrete-only 28/60~GHz experiment. Median path gain is
$-130/-136$~dB and coverage contracts to line-of-sight lobes around the masts.
The explicit concrete-only label avoids extrapolating the ground material beyond
its validated frequency interval.}
\label{fig:mmwave-coverage}
\end{minipage}

\vspace{0.7em}
\begin{minipage}[t]{0.315\textwidth}
\centering
\includegraphics[width=\linewidth]{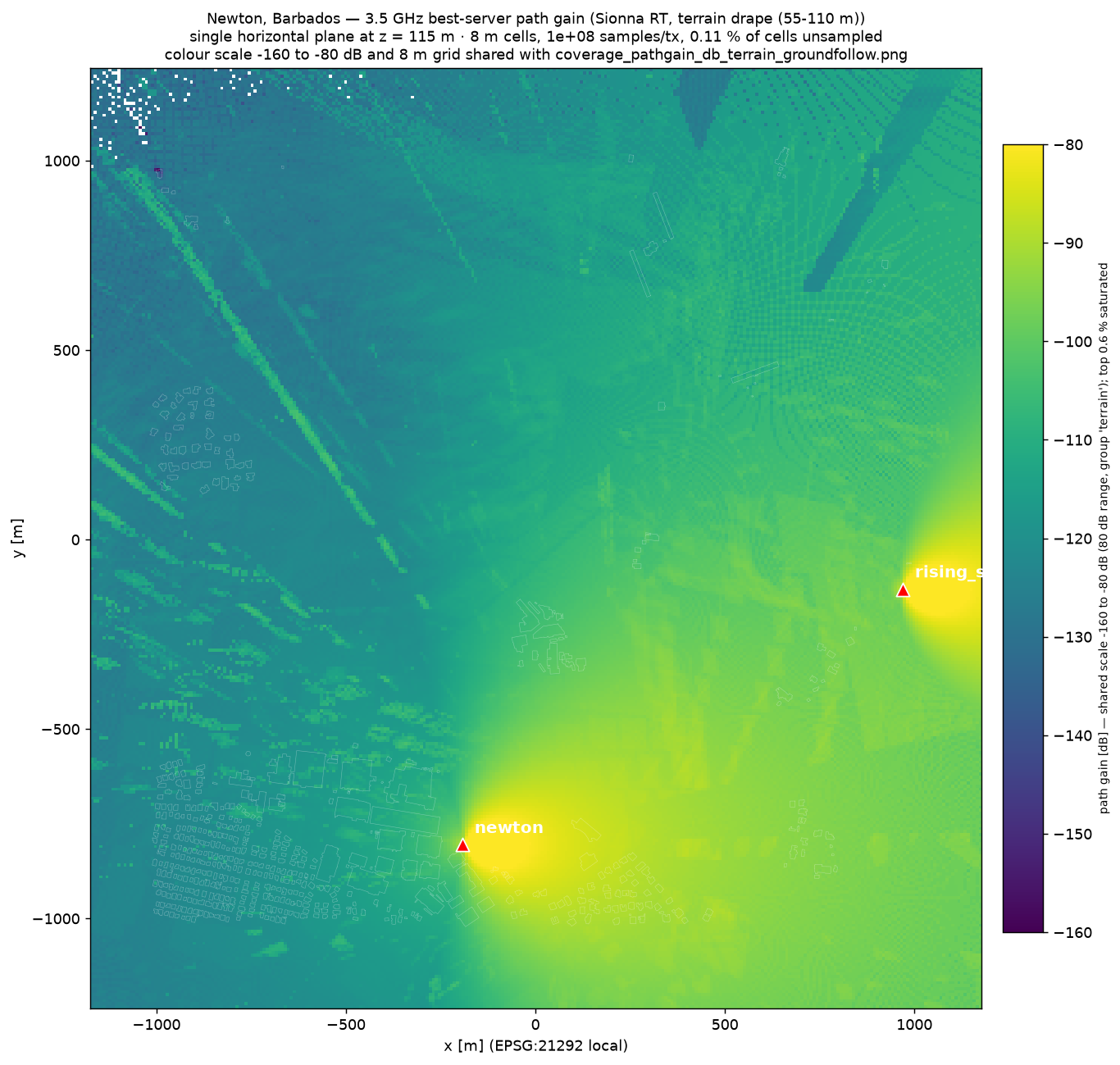}
\captionof{figure}{Conventional 3.5~GHz best-server map over the terrain scene,
sampled on a horizontal receiver plane. Geometry resolves settlement shadowing,
but receiver height above ground varies with relief.}
\label{fig:terrain-plane}
\end{minipage}\hfill
\begin{minipage}[t]{0.315\textwidth}
\centering
\includegraphics[width=\linewidth]{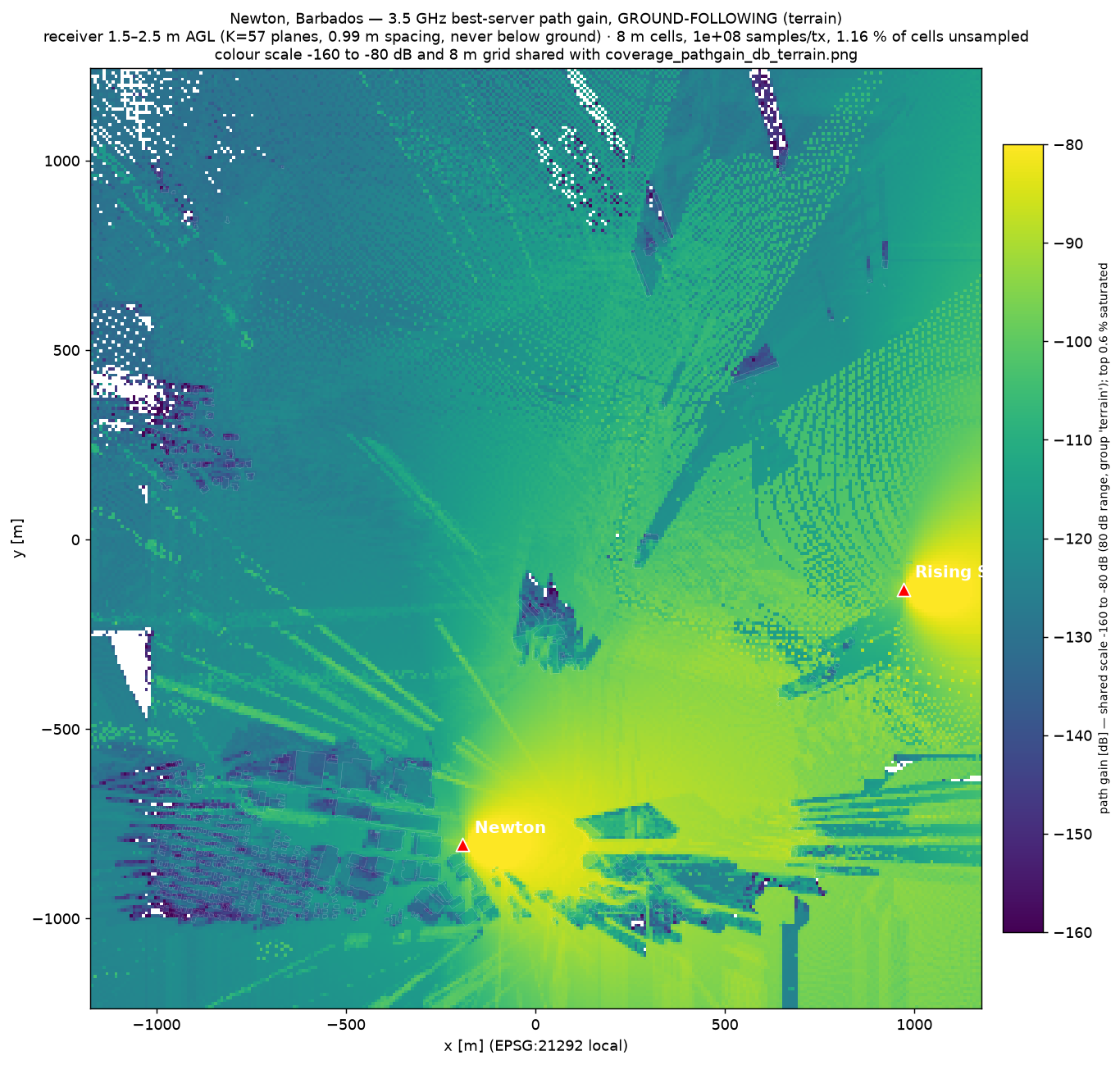}
\captionof{figure}{Ground-following 3.5~GHz map at 1.5~m above local terrain,
assembled from a 57-plane ceiling-rule stack at $10^8$ samples per transmitter, on the
same colour scale and grid as Figure~\ref{fig:terrain-plane}. Residual ridge shadowing appears that the horizontal
map in Figure~\ref{fig:terrain-plane} cannot represent physically.}
\label{fig:ground-following}
\end{minipage}\hfill
\begin{minipage}[t]{0.315\textwidth}
\centering
\includegraphics[width=\linewidth]{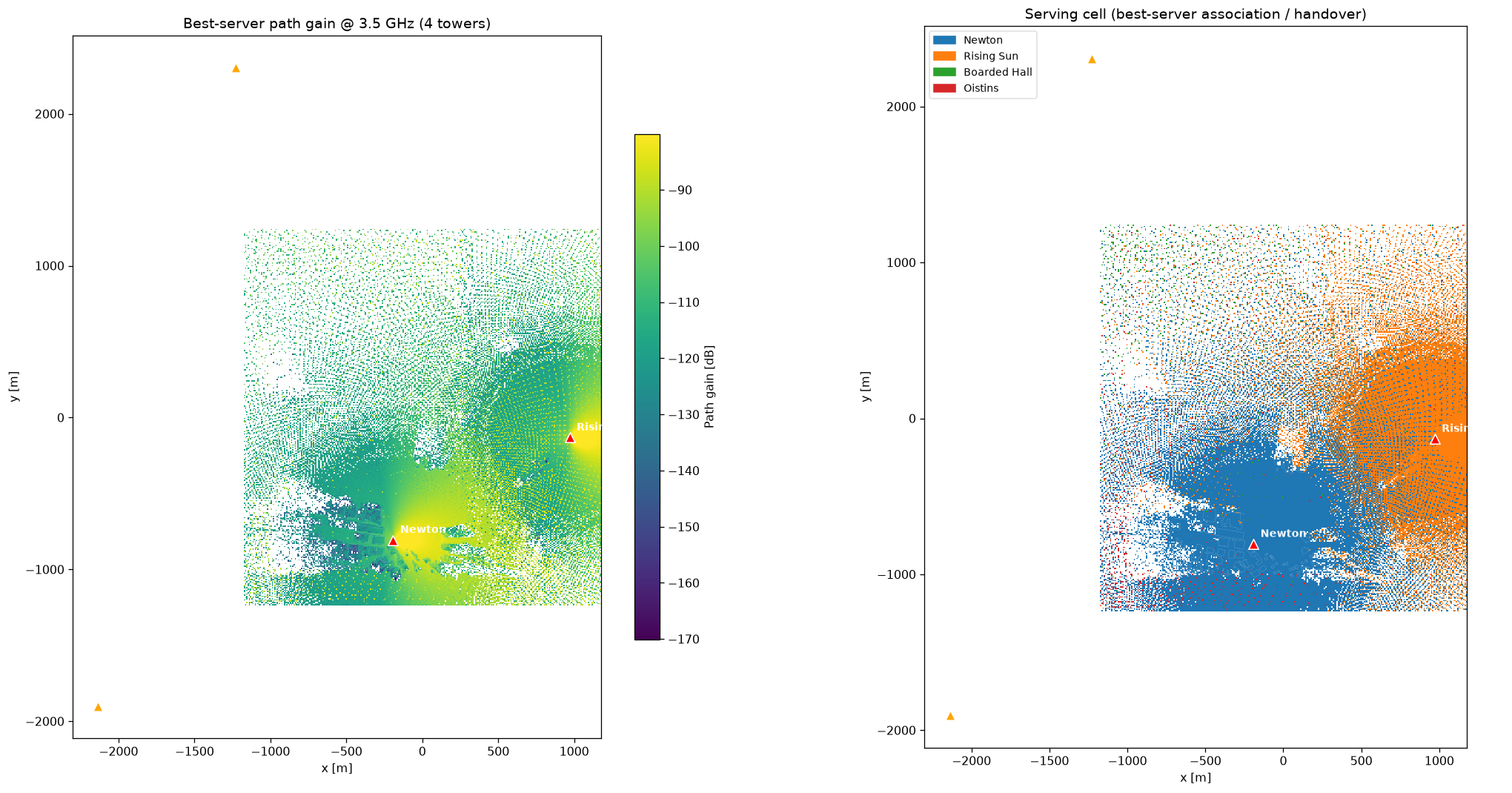}
\captionof{figure}{Four-tower 3.5~GHz best-server gain and serving-cell
association. The deterministic geometry exposes the interference/association
boundary west of Newton and the handover structure used by the live UE-Sim.}
\label{fig:multitower}
\end{minipage}
\end{figure*}

The running Ulap SCOPE twin renders: the island-wide infrastructure network
(130{,}248 buildings, roads, water mains, telecom, ports) draped on 3D terrain; the
same scene in oblique 3D at building fidelity; the authoritative per-parish
vulnerability index as a graded choropleth with a live weather/KPI header; a
government-asset inspect panel (for example Grantley Adams International
Airport (BGI)) that surfaces each feature's source and attributes; and an RF
coverage-planning surface driven by the Sionna propagation engine (receiver
placement, stochastic coverage, and RSS/SINR statistics). The shared Newton
\texttt{scene\_manifest.json} additionally drives live UE-Sim ray-traced coverage
snapshots, connecting the offline deterministic solver to the operational SCOPE
interface. These views exercise built portions of layers 1--4 and 6 of
Table~\ref{tab:map}; screenshots accompany the deployment record in
Figures~\ref{fig:rf-coverage}--\ref{fig:island-overview}, while deterministic
outputs appear in Figures~\ref{fig:freq-sweep}--\ref{fig:multitower}.

\subsection{Roadmap to the Sovereign Cognitive Digital Twin}
\label{sec:roadmap}
The sensing-maturity ladder of \S\ref{sec:physical} is the yardstick, and the roadmap
climbs it while closing the [D] gaps of Table~\ref{tab:map}. \textbf{Phase~0 (built)}
is the ingest-and-serve substrate plus the CPU-only deterministic propagation
workflow and UE-Sim wiring documented above. \textbf{Phase~1 (next)} closes the
highest-value governance gap: content-hash every cube at ingest, stand up the Amini
Chain backbone and register the hashes, and promote the structural checks into an
explicit quality-and-integrity admission gate so outputs become
finance/evacuation-grade. \textbf{Phase~2} replaces the single GeoPackage with a
versioned lakehouse and lifts layers into X/Y/Z/Time cubes for state history and
change detection, while a measurement campaign calibrates the Newton path-gain,
delay, material, and antenna assumptions. \textbf{Phase~3} adds the cognitive
layer---the EKF belief-state synchronization of \S\ref{sec:cognitive}, using the
calibrated deterministic scene as $\mathbf{H}_{\mathrm{background}}$, and ML
surrogates (coastal, hydrology, evacuation) wired to live rainfall fields.
\textbf{Phase~4} adds closed-loop
PPO/AI-RAN orchestration (an O-RAN Near-RT RIC xApp) and the GenAI GovChat query
interface. \textbf{Phase~5} brings the 6G ISAC feed online and enforces the L1/L2/L3
privacy waveforms and cross-layer anomaly detection at the physical layer. The
same workflow transfers by replacing the government source and re-running the
ingest/governance pipeline, from Barbados to the Philippine archipelago.

\section{Conclusion and Future Work}
We have argued that the cheapest path to national-scale climate perception for
island and archipelagic states is not more dedicated sensors but a
re-conception of the network itself as the nation's sensing nervous system,
orchestrated by a sovereign, cognitively closed-loop digital twin. The S-CDT
unifies the built-environment and earth-system digital-twin traditions,
grounds its perception in the emerging ETSI and 3GPP ISAC standards, is designed
to control under realistic RAN delay via belief-state estimation, and treats
sovereignty and physical-layer privacy as primary. The cognitive and
zero-trust layers are specified here but not yet deployed; what is running today
is reported in \S\ref{sec:impl}. The Barbados implementation now adds a
reproducible CPU-only RF scene twin built from sovereign LiDAR-height
buildings, real terrain, and the national tower register. Its frequency,
terrain-following, multi-tower, multipath, and mobility studies instantiate the
site-specific $\mathbf{H}_{\mathrm{background}}$ required by the future ISAC
loop, while the explicit material and calibration limits prevent simulation from
being misreported as deployed 6G sensing. Future work includes: field calibration
of the Newton path-gain and delay predictions; CSI-based rainfall validation
against Caribbean gauge networks; array-aware FR3/FR2 studies; sim-to-real PPO
transfer on an O-RAN testbed; formal verification of the minimized-kernel trust
boundaries; and measurement of the L1/L2/L3 privacy-waveform trade-off curve.

\section{Code Availability}
\label{sec:code-availability}
The open-source software and reproducibility artifacts accompanying this paper
are available at \projectrepository. We release them to the community so that
other small-island and climate-vulnerable states can reproduce these results and
adapt the pipeline to their own national data, which is the practical form the
sovereignty argument of \S\ref{sec:security} takes: a nation cannot hold
sovereign control of a capability it cannot itself run. The release includes the ten-stage
pipeline, pinned environments, manifests, tests, figure scripts,
\texttt{link\_metrics.csv}, the moving-receiver example, and CPU reproduction
instructions. The tagged release corresponding to this paper is \releasetag, archived at
\releasedoi, which records the exact Git commit and solver configuration. Cite
the DOI when reproducing these results: it resolves to that exact tree, whereas
the default branch will move.

\section{Data Availability}
\label{sec:data-availability}
The 19 authoritative static layers underlying the national twin are the property
of the Government of Barbados and were obtained from the Barbados Geoportal,
Lands \& Surveys Department~\cite{bbdgeo}. They are not redistributed with this
paper. The release instead provides acquisition and license instructions,
per-layer checksums and schemas, and non-restricted fixtures sufficient to
exercise the pipeline end to end. Derived products that are ours to
release---the Newton scene manifest, the propagation outputs, and the figure
data---are included in full. Third-party live feeds (Copernicus, ESA WorldCover, OpenSky,
CelesTrak, OSM, OpenCellID, PeeringDB, TeleGeography) remain under their
respective upstream licenses.

\section*{Author Contributions}
Contributions are recorded following the CRediT taxonomy. Each author states their own
roles; none are assigned on another author's behalf. To be completed by the authors before
submission.
All authors read and approved the final manuscript.

\section*{Acknowledgments}
We thank the Government of Barbados for its support of this work: the Ministry
of Industries, Innovation, Science and Technology (MIST), and the Lands \&
Surveys Department for the authoritative national geospatial export that makes
the Barbados twin possible. The interpretations and any errors in this paper are
the authors' own and do not represent the position of the Government of
Barbados.

\sloppy

\fussy

\end{document}